\documentstyle[12pt]{article}

\let\ni=\noindent

\begin{document}

\baselineskip 0.74cm
 
\pagestyle {empty}

\renewcommand{\thefootnote}{\fnsymbol{footnote}}

\newcommand{\CKM}{Cabibbo---Kobayashi---Maskawa }

\newcommand{\UK}{Super--Kamiokande }

\newcommand{\mnu }{\nu_s^{(\mu)} }

\newcommand{\enu }{\nu_s^{(e)} }

~~~

\hfill IFT/99-03
\vspace{0.5cm}

{\large\centerline{\bf Fermion texture and sterile neutrinos}}

\vspace{0.5cm}

{\centerline {\sc Wojciech Kr\'{o}likowski}}

\vspace{0.5cm}

{\centerline {\it Institute of Theoretical Physics, Warsaw University}}

{\centerline {\it Ho\.{z}a 69,~~PL--00--681 Warszawa, ~Poland}}

\vspace{0.5cm}

{\centerline {\bf Abstract}}

\vspace{0.3cm}

\begin{small}

 An explicit form of charged--lepton mass matrix, predicting $ m_\tau = 1776.80
$~MeV  from the experimental values of $ m_e $ and $ m_\mu$ (in good agreement
with the experimental figure $ m_\tau = 1777.05^{+0.29}_{-0.26}$~MeV), is 
applied to three neutrinos $\nu_e $, $\nu_\mu $, $\nu_\tau $ in order to cor\-%
relate tentatively their masses and mixing parameters. It is suggested that for
neutrinos the diagonal elements of the mass matrix are small {\it versus} 
its off-diagonal elements. Under such a conjecture, the neutrino masses, 
lepton \CKM matrix and neutrino oscillation probabilities are calculated in 
the corresponding lowest perturbative order. Then, the nearly maximal mixing 
of $\nu_\mu $ and $\nu_\tau $ is predicted in consistency with the observed 
deficit of atmospheric $\nu_\mu $'s. However, the predicted deficit of solar $
\nu_e $'s is much too small to explain the observed effect, what suggests the 
existence of (at least) one sort, $\enu $, of sterile neutrinos, whose mixing 
with $\nu_e $ would be responsible for the observed deficit. Perspectives for 
applying the same form of mass matrix to quarks are also outlined. Two 
independent predictions of $|V_{ub}|/|V_{cb}| = 0.0753 \pm 0.0032 $ and 
unitary angle $\gamma \simeq 70^\circ $ are deduced from the experimental
values of $|V_{us}|$ and $|V_{cb}|$ (with the use of quark masses $ m_s $, $ 
m_c $ and $ m_b $). In the last three Sections, the option of two sorts, $\enu 
$ and $\mnu $, of sterile neutrinos is considered. They may dominate neutrino 
mixing, and even cause that two extra neutrino mass states (arising then) are 
agents of some tiny neutrino instability and related damping of $\nu_e $ and 
$\nu_\mu $ oscillations. In Appendix, three conventional Majorana sterile 
neutrinos are discussed.

\end{small}

\vspace{0.2cm} 

\ni PACS numbers: 12.15.Ff , 14.60.Pq , 12.15.Hh 
 
\vspace{0.5cm} 

\ni February 1999

\vfill\eject

\pagestyle {plain}

\setcounter{page}{1}

~~~

\vspace{-0.2cm}

\ni {\bf 1. Introduction}

\vspace{0.3cm}

 In this paper, the explicit form of mass matrix invented for three generations
of charged leptons $ e^-\,,\,\mu^-\,$, $\tau^- $, and being surprisingly good
for their masses [1], is applied to three generations of neutrinos $\nu_e $, $
\nu_\mu $, $\nu_\tau $, in order to correlate tentatively their masses and
mixing parameters. This form reads

\vspace{-0.3cm}

\begin{equation}
\left({M}^{(f)}_{\alpha \beta}\right) = \frac{1}{29} \left(\begin{array}{ccc} 
\mu^{(f)}\varepsilon^{(f)} & 2\alpha^{(f)} e^{i\varphi^{(f)}} & 0 \\ & & 
\\ 2\alpha^{(f)} e^{-i\varphi^{(f)}} & 4\mu^{(f)}(80 + \varepsilon^{(f)})/9 
& 8\sqrt{3}\,\alpha^{(f)} e^{i\varphi^{(f)}} \\ & & 
\\ 0 & 8\sqrt{3}\,\alpha^{(f)} e^{-i\varphi^{(f)}} & 24\mu^{(f)}
(624 + \varepsilon^{(f)})/25 \end{array}\right) \;,
\end{equation}

\vspace{-0.1cm}

\ni where the label $f = \nu,\,e $ denotes neutrinos and charged leptons, 
respectively, while $\mu^{(f)}$, $\varepsilon^{(f)}$, $\alpha^{(f)}$ and 
$\varphi^{(f)}$ are real constants to be determined from the present and 
future experimental data for lepton masses and mixing parameters ($\mu^{(f)}$
and $\alpha^{(f)}$ are mass--dimensional). In our approach, neutrinos are 
assumed to carry pure Dirac masses.

 Here, the form (1) of mass matrices $\left({M}^{(\nu)}_{\alpha \beta}\right)$ 
and $\left({M}^{(e)}_{\alpha \beta}\right)$ may be considered as a detailed 
ansatz to be compared with the lepton data. However, in the past, we have 
presented an argument [2,1] in favour of the form (1), based on: ({\it i}) 
K\"{a}hler--like generalized Dirac equations (interacting with the Standard 
Model gauge bosons) whose {\it a priori} infinite series is necessarily reduced
(in the case of fermions) to three Dirac equations, due to an intrinsic Pauli 
principle, and ({\it ii}) an ansatz for the fermion mass matrix, suggested by 
the above three--generation characteristics ({\it i}).

 In the case of charged leptons, assuming that the off--diagonal elements of 
the mass matrix $\left({M}^{(e)}_{\alpha \beta}\right)$ can be treated as a 
small perturbation of its diagonal terms ({\it i.e.}, that $\alpha^{(e)}/\mu^
{(e)}$ is small enough), we calculate in the lowest perturbative order [1]

\vspace{-0.2cm}

\begin{eqnarray}
m_\tau & = & \left[ 1776.80 + 10.2112 \left(\frac{\alpha^{(e)}}{\mu^{(e)}}
\right)^2\,\right]\;{\rm MeV} \;, \nonumber \\
\mu^{(e)} & = & 85.9924\;{\rm MeV} + O\left[\left(\frac{\alpha^{(e)}}{\mu^{(e)}
}\right)^2\,\right]\,\mu^{(e)} \;,\nonumber \\
\varepsilon^{(e)} & = & 0.172329 + O\left[\left(\frac{\alpha^{(e)}}{\mu^{(e)
}}\right)^2 \right]\;,
\end{eqnarray}

\vspace{-0.1cm}

\ni when the experimental values of $ m_e $ and $ m_\mu $ [3] are used as 
inputs. In Eqs. (2), the first terms are given as $\stackrel{\circ}{m}_\tau =
6(351m_\mu - 136 m_e)/125 $, $\stackrel{\circ}{\mu}^{(e)} = 29(9m_\mu - 4 m_e
)/320 $ and $\stackrel{\circ}{\varepsilon}^{(e)} = 320m_e/(9m_\mu - 4 m_e)$, 
respectively. We can see that the predicted value of $ m_\tau $ agrees very 
well with its experimental figure $ m_\tau^{\rm exp} = 1777.05^{+0.29}_{-0.26}
$~MeV [3], even in the zero perturbative order. To estimate $\left(\alpha^{(e)}
/\mu^{(e)}\right)^2 $, we can take this experimental figure as another input, 
obtaining


\begin{equation}
\left(\frac{\alpha^{(e)}}{\mu^{(e)}}\right)^2 = 0.024^{+0.028}_{-0.025} \;,
\end{equation}

\vspace{0.1cm}

\ni  which value is not inconsistent with zero. Hence, $\alpha^{(e)\,2} = 180
^{+210}_{-190}\,{\rm MeV}^2 $ due to Eq. (2).

 For the unitary matrix $\left({U}^{(e)}_{\alpha \beta}\right)$, diagonalizing the 
charged--lepton mass matrix $\left({M}^{(e)}_{\alpha \beta}\right)$ according to the 
relation $ U^{(e)\,\dagger}\,M^{(e)}\,U^{(e)} = {\rm diag}(m_e\,,\,m_\mu\,,\,
m_\tau)$, we get in the lowest perturbative order 


\begin{equation}
\left( U^{(e)}_{\alpha \beta}\right) = \left(\begin{array}{ccc} 1 - \frac{2}{29^2}\left(
\frac{\alpha^{(e)}}{m_\mu}\right)^2 & \frac{2}{29} \frac{\alpha^{(e)}}{m_\mu}
e^{i\varphi^{(e)}} & \frac{16\sqrt{3}}{29^2} \left(\frac{\alpha^{(e)}}{m_\tau} 
\right)^2 e^{2i \varphi^{(e)}} \\ & & \\ -\frac{2}{29}\frac{\alpha^{(e)}}
{m_\mu} e^{-i\varphi^{(e)}} & 1 - \frac{2}{29^2}\left(\frac{\alpha^{(e)}}{
m_\mu}\right)^2 - \frac{96}{29^2}\left(\frac{\alpha^{(e)}}{m_\tau}\right)^2 & 
\frac{8\sqrt{3}}{29}\frac{\alpha^{(e)}}{m_\tau}e^{i\varphi^{(e)}} \\  & & \\
\frac{16\sqrt{3}}{29^2}\frac{\alpha^{(e)\,2}}{m_\mu\,m_\tau}\,e^{-
2i \varphi^{(e)}} & -\frac{8\sqrt{3}}{29}\frac{\alpha^{(e)}}{m_\tau} e^{-i
\varphi^{(e)}} & 1 - \frac{96}{29^2}\left(\frac{\alpha^{(e)}}{m_\tau}\right)^2 
\end{array} \right) \;.
\end{equation}

\vspace{0.3cm}

\ni {\bf 2. Neutrino masses and mixing parameters}

\vspace{0.3cm}

 In the case of neutrinos, because of their expected tiny mass scale $\mu^{(\nu
)}$, we will tentatively conjecture that the diagonal elements of the mass 
matrix $\left( M^{(\nu)}_{\alpha \beta}\right)$ can be treated as a small perturbation of
its off--diagonal terms ({\it i.e.}, that $\mu^{(\nu)}/\alpha^{(\nu)}$ is small
enough). In addition, we put $\varepsilon^{(\nu)} = 0 $ {\it i.e.}, $ M^{(
\nu)}_{11} = 0 $. Then, we calculate in the lowest perturbative order the 
following neutrino masses:


\begin{eqnarray}
m_{\nu_1} & = & \frac{|M^{(\nu)}_{12}|^2\,M^{(\nu)}_{33}}{|M^{(\nu)}_{12}|^2
+ |M^{(\nu)}_{23}|^2} = \frac{1}{49}M^{(\nu)}_{33}  = \frac{1}{49} \xi |M^{(\nu
)}_{12}|\;, \nonumber \\ 
m_{\nu_2,\,\nu_3} & = &  \mp \sqrt{|M^{(\nu)}_{12}|^2 + |M^{(\nu)}_{23}|^2}
+ \frac{1}{2}\left( \frac{48}{49}M^{(\nu)}_{33} + M^{(\nu)}_{22}\right)
\nonumber \\ & = & \left[ \mp 7 + \frac{1}{2}\left( \frac{48}{49} \xi + \chi
\right)\right] |M^{(\nu)}_{12}| \;,
\end{eqnarray}


\ni where


\begin{eqnarray}
\xi & \equiv & \frac{M^{(\nu)}_{33}}{|M^{(\nu)}_{12}|}= \frac{7488}{25}\frac{
\mu^{(\nu)}}{\alpha^{(\nu)}} = 299.52\frac{\mu^{(\nu)}}{\alpha^{(\nu)}}
\;, \nonumber \\ 
\chi & \equiv & \frac{M^{(\nu)}_{22}}{|M^{(\nu)}_{12}|}= \frac{160}{9}\frac{
\mu^{(\nu)}}{\alpha^{(\nu)}} = \frac{125}{2106} \xi = \frac{1}{16.848} \xi \;,
\end{eqnarray}

\vspace{0.1cm}

\ni are relatively small by our perturbative conjecture, while


\begin{equation}
|M^{(\nu)}_{12}| = \frac{2}{29}\alpha^{(\nu)}\;,\; |M^{(\nu)}_{23}| = 
\frac{8\sqrt{3}}{29}\alpha^{(\nu)} =\sqrt{48} |M^{(\nu)}_{12}| \;.
\end{equation}

\vspace{0.1cm}

\ni As seen from Eqs. (5), the actual perturbative parameters are not $\xi $ 
and $\chi $, but rather $\xi/7 $ and $\chi/7 $, what is confirmed later in Eqs.
(9). Note that $ m_{\nu_2} < 0 $, the minus sign being irrelevant in the 
relativistic case, where only $ m_{\nu_2}^2 $ is measured ({\it cf.} Dirac 
equation): $ |m_{\nu_2}| $ may be considered as a phenomenological mass of $
\nu_2 $.

 Using Eqs. (5), we can write the formula


\begin{equation}
m_{\nu_3}^2 - m_{\nu_2}^2 = 14\left(\frac{48}{49}\xi + \chi\right)|M^{(\nu)}_{
12}|^2 = 20.721\, \alpha^{(\nu)} \mu^{(\nu)} \;,
\end{equation}

\vspace{0.1cm}

\ni which will enable us to determine the product  $\alpha^{(\nu)}\mu^{(\nu)}$ 
from the observed deficit of atmospheric neutrinos $\nu_\mu $, if $\nu_\mu 
\rightarrow \nu_\tau $ oscillations are really responsible for this effect.

 We calculate also the unitary matrix $\left(U_{\alpha\,i}^{(\nu)}\right)$ dia%
gonal\-izing  the neutrino mass matrix $\left(M_{\alpha \beta}^{(\nu)}\right)$ 
according to the relation $ U^{(\nu)\,\dagger} M^{(\nu)} U^{(\nu)} = {\rm diag}
(m_{\nu_1}\,,\,m_{\nu_2}\,,\,m_{\nu_3})$. In the lowest perturbative order we 
obtain

\vspace{-0.2cm}

\begin{eqnarray}
U^{(\nu)}_{11} & = & \sqrt{ \frac{48}{49} }\left[1 - \left(\frac{24}{49^3} -
\frac{1}{49^4}\right)\,\xi^2\right] \;, \nonumber \\
U^{(\nu)}_{21} & = & \frac{1}{49}\sqrt{\frac{48}{49}} \xi e^{-i\varphi^{(\nu)}
} \;, \nonumber \\
U^{(\nu)}_{31} & = & -\frac{1}{7}\left[1 - \left(\frac{73}{49^3} - 
\frac{1}{49^4}\right) \xi^2 + \frac{1}{49}\xi \chi \right] e^{-2i 
\varphi^{(\nu)}} \;, \nonumber \\
U^{(\nu)}_{12} & = & -\frac{1}{\sqrt{2}} \frac{1}{7} \left(1 +
\frac{36}{7\cdot 49} \xi + \frac{1}{28}\chi \right)\,e^{i\varphi^{(\nu)}}
\;, \nonumber \\
U^{(\nu)}_{22} & = & \frac{1}{\sqrt{2}} \left(1 + \frac{12}{7\cdot 49} \xi 
- \frac{1}{28}\chi \right) \;, \nonumber \\
U^{(\nu)}_{32} & = & -\frac{1}{\sqrt{2}} \sqrt{\frac{48}{49}} \left(1 -
\frac{13}{7\cdot 49} \xi + \frac{1}{28} \chi \right)\,e^{-i \varphi^{(\nu)}}
\;, \nonumber \\
U^{(\nu)}_{13} & = & \frac{1}{\sqrt{2}} \frac{1}{7} \left(1 -
\frac{36}{7\cdot 49} \xi - \frac{1}{28}\chi \right)\,e^{2i \varphi^{(\nu)}}
\;, \nonumber \\
U^{(\nu)}_{23} & = & \frac{1}{\sqrt{2}} \left(1 - \frac{12}{7\cdot 49} \xi + 
\frac{1}{28}\chi \right)\,e^{i\varphi^{(\nu)}} \;, \nonumber \\
U^{(\nu)}_{33} & = & \frac{1}{\sqrt{2}} \sqrt{\frac{48}{49}} \left(1 +
\frac{13}{7\cdot 49} \xi - \frac{1}{28} \chi \right) 
\end{eqnarray}

\vspace{-0.1cm}

\ni with $\chi = (125/2106)\xi = \xi/16.848 $.

 Denoting by $\nu_\alpha = \nu_e\,,\,\nu_\mu \,,\,\nu_\tau $ and $\nu_i = 
\nu_1\,,\,\nu_2\,,\,\nu_3 $ the flavor and mass neutrino fields, respectively, 
we have the unitary transformation

\vspace{-0.2cm}

\begin{equation}
\nu_\alpha = \sum_i \left( V^\dagger \right)_{\alpha\,i}\nu_i = \sum_i V^*_{i\,
\alpha} \nu_i \;,
\end{equation}

\vspace{-0.1cm}

\ni where the lepton counterpart $\left( V_{i\,\alpha}\right) $ of the \CKM 
matrix is given as $ V = U^{(\nu)\,\dagger} U^{(e)} \simeq U^{(\nu)\,\dagger}$ 
or 

\vspace{-0.2cm}

\begin{equation}
V_{i\,\alpha} = \sum_\beta \left( U^{(\nu)\,\dagger} \right)_{i\,\beta}
U^{(e)}_{\beta \alpha} \simeq U^{(\nu)\,*}_{\alpha\,i} \;,
\end{equation}

\vspace{-0.1cm}

\ni the approximate equality being valid for negligible $\alpha^{(e)}/\mu^{(e)}
$ when $ U_{\beta \alpha}^{(e)} \simeq \delta_{\beta \alpha}$ due to Eq. (4).
Of course, in Eqs. (9) we wrote $\alpha = 1,2,3 $ for simplicity. From Eq. 
(10), we get the unitary transformation $ |\nu_\alpha\rangle = \sum_i |\nu_i
\rangle V_{i\,\alpha}$, where $ |\nu_\alpha\rangle = \nu^\dagger_\alpha |0 
\rangle $ and $ |\nu_i\rangle = \nu^\dagger_i |0 \rangle $ are flavor and mass 
neutrino states{\footnote{In place of $\nu_i = \sum_\alpha V_{i\,\alpha} 
\nu_\alpha $ one might use the notation $\nu'_\alpha = \sum_\beta V_{\alpha\,
\beta} \nu_\beta $, analoguous to $ d'_\alpha = \sum_\beta V_{\alpha \beta} 
d_\beta $ customary in the case of quarks where $ V_{\alpha \beta} = \sum
_\gamma\left( U^{(u)\,\dagger}\right)_{\alpha\,\gamma} U^{(d)}_{\gamma\,\beta}
$.}}.

 In the limit of $\mu^{(\nu)} \rightarrow 0 $ (implying $\xi \rightarrow 0 $ 
and $\chi \rightarrow 0 $), we obtain from Eqs. (10), (11) and (9) the follow%
ing unperturbed mixing formulae for $\nu_1\,,\,\nu_2\,,\,\nu_3 $ :

\vspace{-0.2cm}

\begin{eqnarray}
\nu_e  & \rightarrow & \frac{1}{7} \left[\sqrt{48} \nu_1 e^{-i \varphi^{(\nu)
}} - \frac{1}{\sqrt{2}}\left(\nu_2 - \nu_3 e^{i \varphi^{(\nu)}}\right)\right]
e^{i \varphi^{(\nu)}} \;, \nonumber \\ 
\nu_\mu & \rightarrow & \frac{1}{\sqrt{2}}\left(\nu_2 + \nu_3 e^{i \varphi^{(
\nu)}} \right) \;, \nonumber \\ 
\nu_\tau & \rightarrow &  -\frac{1}{7} \left[\nu_1 e^{-i \varphi^{(\nu)}} +
\sqrt{\frac{48}{2}}\left(\nu_2 - \nu_3 e^{i \varphi^{(\nu)}}\right)\right]
e^{-i \varphi^{(\nu)}} \;.
\end{eqnarray}

\vspace{-0.1cm}

\ni These display the maximal mixing between $\nu_2 $ and $\nu_3 $ in all three
cases and a smaller mixing of $\left[\nu_2 - \nu_3 \exp \left(i\varphi^{(\nu)}
\right)\right]/\sqrt{2}$ with $\nu_1 $ in the cases of $\nu_e$ and $\nu_\tau $, 
giving a minor admixture to $\nu_e $ and a dominating admixture to $\nu_\tau $ 
(in $\nu_\mu $ there is no admixture of $\nu_1 $).


\vspace{0.3cm}

\ni {\bf 3. Neutrino oscillations}

\vspace{0.3cm}

 Once knowing the elements $ V_{i\,\alpha}$ of the lepton \CKM matrix, we can 
calculate the probabilities of neutrino oscillations $\nu_\alpha \rightarrow 
\nu_\beta$ (in the vacuum) making use of the general formula

\vspace{0.1cm}

\begin{equation}
P(\nu_\alpha \rightarrow \nu_\beta) = |\langle\nu_\beta |\nu_\alpha(t)\rangle
|^2 = \sum_{i\,j}V_{j\,\beta}V^*_{j\,\alpha}V^*_{i\,\beta}V_{i\,\alpha} 
e^{2i x_{j\,i}} \;,
\end{equation}

\vspace{0.1cm}

\ni where $|\nu_\alpha(t)\rangle = \exp(-i H t) |\nu_\alpha \rangle $ and 


\begin{equation}
x_{j\,i} = 1.26693\, \Delta m^2_{j\,i}\, L/E \;\;,\;\;\Delta m^2_{j\,i} = 
m^2_{\nu_j} - m^2_{\nu_i}\;\;,
\end{equation}


\ni if $\Delta m^2_{j\,i}$, $ L $ and $ E $ are measured in eV$ ^2 $, km and 
GeV, respectively, with $ L = t $ and $ E = |\vec{p}|\;\;(c = 1 = \hbar) $ 
denoting the experimental baseline and neutrino energy.

 It is not difficult to show that for the mass matrix $\left( M^{(\nu)}_{\alpha \beta}
\right)$, as it is given in Eq. (1), the quartic products of $V_{i\,\alpha}$'s 
in Eq. (13) are always real (for any phase $\varphi^{(\nu)}$), if only $V_{i\,
\alpha} = U^{(\nu)\,*}_{\alpha\,i}$ ({\it i.e.}, $ U^{(e)}_{\beta \alpha} = 
\delta_{\beta \alpha}$). This implies that $P(\nu_\alpha \rightarrow \nu_\beta
) = P(\nu_\beta \rightarrow \nu_\alpha)$. In general, the last relation is valid
in the case of CP invariance which, under the CPT theorem, provides the time--%
reversal invariance. Because of the real values of quartic products of $V_{i\,
\alpha}$'s, the formula (13) can be rewritten as


\begin{equation}
P(\nu_\alpha \rightarrow \nu_\beta) = \delta_{\beta \alpha} - 4 \sum_{i<j}
V_{j\,\beta}V^*_{j\,\alpha}V^*_{i\,\beta}V_{i\,\alpha} \sin^2 x_{j\,i} 
\end{equation}


\ni without the necessity of introducing phases of these products.

 With the lowest--order perturbative expressions (9) for $V_{i\,\alpha} = U^{(
\nu)\,*}_{\alpha\,i}$, the formula (15) leads to the following forms of 
appearance oscillation probabilities:


\begin{eqnarray}
\lefteqn{P\left(\nu_\mu \rightarrow \nu_e  \right) = \frac{1}{49}\sin^2 x_{32}}
\nonumber \\ &  & + \frac{96}{7\cdot 49^2} \xi \left[\left(1 + \frac{48}{7
\cdot 49} \xi \right) \sin^2 x_{21} - \left(1 - \frac{48}{7\cdot 49} \xi\right)
\sin^2 x_{31}\right] \;,\\
\lefteqn{P\left(\nu_\mu \rightarrow \nu_\tau\right) = \frac{48}{49}\sin^2 x_{32
}} \nonumber \\ &  & + \frac{96}{7\cdot 49^2} \xi \left[ -\left(1 - \frac{1}{
7\cdot 49} \xi \right) \sin^2 x_{21} + \left(1 + \frac{1}{7\cdot 49} \xi\right)
\sin^2 x_{31}\right] \;,\\
\lefteqn{P\left(\nu_e \rightarrow \nu_\tau  \right) = - \frac{48}{49^2}
\sin^2 x_{32}} \nonumber \\ &  & + \frac{96}{49^2}\left[\left(1 + \frac{23}{7
\cdot 49} \xi + \frac{1}{14} \chi \right) \sin^2 x_{21} + \left(1 - \frac{23
}{7\cdot 49} \xi - \frac{1}{14} \chi \right) \sin^2 x_{31}\right] 
\end{eqnarray}


\ni as well as of survival oscillation probabilities :


\begin{eqnarray}
\lefteqn{P\left(\nu_e \rightarrow \nu_e  \right) = 1 - \frac{1}{49^2}
\sin^2 x_{32}} \nonumber \\ &  & - \frac{96}{49^2}\left[\left(1 + \frac{72}{7
\cdot 49} \xi + \frac{1}{14} \chi\right) \sin^2 x_{21} + \left(1 - \frac{72}{7
\cdot 49} \xi - \frac{1}{14} \chi\right)\sin^2 x_{31}\right] \;,\\
\lefteqn{P\left(\nu_\mu \rightarrow \nu_\mu\right) = 1 - \sin^2 x_{32}
- \frac{96}{49^3} \xi^2 \left(\sin^2 x_{21} + \sin^2 x_{31}\right) \;,} \\
\lefteqn{P\left(\nu_\tau \rightarrow \nu_\tau\right) = 1 - \left(\frac{48}{49}
\right)^2\sin^2 x_{32}} \nonumber \\ &  & - \frac{96}{49^2}\left[\left(1 -
\frac{26}{7\cdot 49} \xi + \frac{1}{14} \chi \right) \sin^2 x_{21} + \left(
1 + \frac{26}{7\cdot 49} \xi - \frac{1}{14} \chi \right) \sin^2 x_{31}\right] 
\;.
\end{eqnarray}

\ni Thus, we get $ P\left(\nu_e \rightarrow \nu_e \right) + P\left(\nu_e 
\rightarrow \nu_\mu\right) + P\left(\nu_e \rightarrow \nu_\tau\right) = 1 $ and
two other obvious summation rules for probabilities. Among these probabilities,
$P\left(\nu_\mu \rightarrow \nu_\mu\right)$ displays (in the lowest perturbat%
ive order) maximal mixing between $\nu_2 $ and $\nu_3 $.

 In the lowest perturbative order,


\begin{equation}
x_{31} - x_{21} = x_{32} = 14\left(\frac{48}{49} \xi + \chi \right)
\left(1.26693 |M^{(\nu)}_{12}|\, L/E \right)
\end{equation}

\ni due to Eqs. (8) and (14). Hence,


\begin{equation}
\sin^2 x_{31} = \sin^2 x_{21} + x_{32}\sin 2x_{21} + x^2_{32}\sin 2x_{21}
\end{equation}

\ni in experiments where $ x_{32} \ll \pi/2 $. When in such cases the relation
(23) is inserted into the formulae (16), (17) and (20), its $ x_{32}$ and 
$ x^2_{32}$ terms can be neglected in the lowest perturbative order.

 Note that the mass formulae (5) imply $m^2_{\nu_1} \ll m^2_{\nu_2} \stackrel{
_<}{_\sim} m^2_{\nu_3} $, where $m^2_{\nu_1}/m^2_{\nu_2,\nu_3} = \xi^2/49^3 
+ O(\xi^3)$ and $m^2_{\nu_2}/m^2_{\nu_3} = 1 -(2/7)(48 \xi/49 + \chi) + O(
\xi^3)$. Thus, the inequality $x_{31} \stackrel{_>}{_\sim} x_{21} \gg x_{32}
$ holds in all neutrino oscillation experiments (with some given $ L $ and 
$ E $).

 We have calculated the neutrino masses, lepton \CKM matrix and neutrino oscil%
lation probabilities also in the next to lowest perturbative order. Then, in 
Eqs. (5) the mass $m_{\nu_1}$ gets no quadratic correction, while $m_{\nu_2}$ 
and $m_{\nu_3}$ are corrected by the terms

\begin{equation}
\mp \frac{1}{14}\left(\frac{13\cdot 48}{49^2} \xi^2 - \frac{24}{49} \xi \chi
+ \frac{1}{4} \chi^2\right)|M^{(\nu)}_{12}| \;,
\end{equation}

\ni respectively. Among the derived oscillation formulae, Eq. (20), for 
instance, is extended to the form

\vspace{-0.2cm}

\begin{eqnarray}
\lefteqn{P\left(\nu_\mu \rightarrow \nu_\mu\right) = 1 - \left(1 - 
\frac{672}{49^3} \xi^2 + \frac{24}{49^2} \xi \chi - \frac{1}{4\cdot 49} 
\chi^2 \right) \sin^2 x_{32} } \nonumber \\ & & \;\;\;\;\;\;\;\;\;\;\;
\;\;\;\;\;\;\;\;\;\; - \frac{96}{49^3} \xi^2 \left( \sin^2 x_{21} + 
\sin^2 x_{31}\right) \nonumber \\ 
& &\;\;\;\;\;\;\;\;\;\;\;\;\;\; = 1 - \left(1 - 0.00514\,\xi^2 \right)\sin^2 
x_{32} - 0.000816\,\xi^2 \left(\sin^2 x_{21} + \sin^2 x_{31}\right)\;\;\;\;\;
\end{eqnarray}

\vspace{-0.1cm}

\ni displaying nearly maximal mixing between $\nu_2 $ and $\nu_3 $. 

In the case of \UK atmospheric neutrino experiment [4], if $\nu_\mu \rightarrow
\nu_\tau $ oscillations are responsible for the observed deficit of atmospheric
$\nu_\mu $'s, we have $ x_{\rm atm} = x_{32} \ll x_{21} \stackrel{_<}{_\sim} 
x_{31}$, what implies that $\sin^2 x_{21} = \sin^2 x_{31} = 1/2 $ due to 
averaging over many oscillation lengths. Then, Eq. (25) leads to the following 
effective two--flavor oscillation formula:

\vspace{-0.2cm}

\begin{equation}
P\left(\nu_\mu \rightarrow \nu_\mu \right) = 1 - \left(1 - 0.00350\,\xi^2
\right) \sin^2 x_{32}\;,
\end{equation}

\ni if we assume in Eq. (25) that $0.000816 \xi^2 = 0.000816 \xi^2 (2\sin^2 
x_{32})$ effectively. Identifying the estimation (26) with the two--flavor formula
fitted in the \UK experiment, we obtain the limits

\vspace{-0.2cm}

\begin{eqnarray}
1 - 0.00350\,\xi^2 & \equiv & \sin^2 2\theta_{\rm atm} \sim 0.82\;\;{\rm to}
\;\; 1 \;, \nonumber \\ \Delta m_{32}^2 & \equiv & \Delta m_{\rm atm}^2 \sim 
(0.5\;\;{\rm to}\;\;6)\times 10^{-3}\,{\rm eV}^2 \;.
\end{eqnarray}

\ni Hence, $\xi \sim 7.17$ to 0 and

\vspace{-0.2cm}

\begin{eqnarray}
\frac{\mu^{(\nu)}}{\alpha^{(\nu)}} & \equiv & 0.00334 \xi \sim 0.0239\;\;{\rm 
to}\;\;0 \;, \nonumber \\ \alpha^{(\nu)}\mu^{(\nu)} & \equiv & 0.483\Delta m_{32
}^2 \sim (0.241\;\;{\rm to}\;\;2.90)\times 10^{-4}\,{\rm eV}^2 \;,
\end{eqnarray}

\ni where Eqs. (6) and (8) are used. For instance, with $\sin^2 2\theta_{\rm 
atm} \sim 0.999 $ and $\Delta m^2_{\rm atm} \sim 5\times 10^{-3}\,{\rm eV}^2 $,
we get $\xi \sim 0.535 $ and

\begin{equation}
\frac{\mu^{(\nu)}}{\alpha^{(\nu)}} \sim 0.00178 \;\;,\;\; \alpha^{(\nu)}
\mu^{(\nu)} \sim 2.41\times 10^{-4}\,{\rm eV}^2 \;,
\end{equation}

\ni what gives the estimation

\vspace{-0.2cm}

\begin{equation}
\alpha^{(\nu)} \sim 0.368\,{\rm eV}\;\;,\;\;\mu^{(\nu)} \sim 6.55\times 10^{-4}
\,{\rm eV} \;.
\end{equation}

\ni Note that $ \xi < 1 $ for $\sin^2 2\theta_{\rm atm} > 0.9965 $. As was 
already mentioned, our actual perturbative parameters are not $\xi $ and $\chi 
$, but rather $\xi/7 $ and $\chi/7 = 0.0594 \xi/7 $.

 Having estimated $\alpha^{(\nu)}$ and $\mu^{(\nu)}$, we can calculate neutrino
masses from Eqs. (5) with (6) and (7). Making use of the values (30) (valid for
$\sin^2 2\theta_{\rm atm} \sim 0.999 $ and $\Delta m^2_{\rm atm} \sim 5\times 
10^{-3}\,{\rm eV}^2 $), we obtain

\vspace{-0.2cm}

\begin{equation}
m_{\nu_1} \sim 2.76\times 10^{-4}\,{\rm eV} \;\;,\;\;m_{\nu_2} \sim -1.71
\times 10^{-1}\,{\rm eV} \;\;,\;\;m_{\nu_3} \sim 1.85\times 10^{-1}\,{\rm eV}.
\end{equation}

\ni Because of the smallness of these masses, the neutrinos $\nu_1 $, $\nu_2 $,
$\nu_3 $ are not likely to be responsible for the entire hot dark matter.

 In the case of solar neutrino experiments, all three popular fits [5] of the 
observed deficit of solar $\nu_e $'s to an effective two--flavor oscillation 
formula require $\Delta m^2_{\rm sol} \ll \Delta m^2_{\rm atm}$ what implies $
\Delta m^2_{\rm sol}\ll \Delta m^2_{32} \ll \Delta m^2_{21} \stackrel{_<}{_\sim}
\Delta m^2_{31}$, if $\nu_\mu \rightarrow \nu_\tau $ oscillations are res\-%
ponsible for the deficit of atmospheric $\nu_\mu $'s. Then, $ x_{\rm sol} \ll 
x_{32} \ll x_{21} \stackrel{_<}{_\sim} x_{31}$, giving $\sin^2 x_{32} = \sin^2 
x_{21} = \sin^2 x_{31} = 1/2 $ due to averaging over many oscillation lengths.
In such a case, Eq. (19) leads to

\vspace{-0.2cm}

\begin{equation}
P\left(\nu_e \rightarrow \nu_e \right) = 1 - \frac{193}{2\cdot 49^2} = 1 -
0.0402 = 0.960 \;,
\end{equation}

\ni predicting only a 4\% deficit of solar $\nu_e$'s, much too small to explain
solar neutrino observations.

 An intriguing situation arises in the case of formula (16) for $ P\left(
\nu_\mu \rightarrow \nu_e \right)$, if $\nu_\mu \rightarrow \nu_\tau $ oscil%
lations really cause the bulk of deficit of atmospheric $\nu_\mu $'s. Then, 
for a new $ x_{\rm new} = x_{32} \ll x_{21} \stackrel{_<}{_\sim} x_{31}$ (with 
some new $ L $ and $ E $) we may have $\sin^2 x_{21} = \sin^2 x_{31} = 1/2$ due
to averaging over many oscillation lengths and so, infer from Eq. (16) that

\vspace{-0.2cm}

\begin{equation}
P\left(\nu_\mu \rightarrow \nu_e \right) = \frac{1}{49} \sin^2 x_{32} + \frac{2
\cdot 48^2}{49^4} \xi^2 \sim 0.0204 \sin^2 x_{32} + 2.29\times 10^{-4}\;,
\end{equation}

\ni where $\xi^2 \sim 0.286 $ (what is valid for $\sin^2 2\theta_{\rm atm} \sim
0.999 $ and $\Delta m^2_{\rm atm} \sim 5\times 10^{-3}\,{\rm eV}^2 $). Such a 
predicted oscillation amplitude $\sin^2 2\theta_{\rm new} \sim 0.02 $ would lie
in the range of $\sin^2 2\theta_{\rm LSND}$ estimated in the positive (though 
still requiring confirmation) LSND accelerator experiment on $\nu_\mu 
\rightarrow \nu_e $ oscillations [6]. However, the lower limit $\Delta m^2_{
\rm LSND} \stackrel{_>}{_\sim} 0.1\,{\rm eV}^2 $ reported by this experiment 
is by one order of magnitude larger than the \UK upper limit $\Delta
m^2_{32} \stackrel{_<}{_\sim} 0.01\,{\rm eV}^2 $. On the other hand, the small 
predicted oscillation amplitude $\sin^2 2\theta_{\rm new} \sim 0.02 $ would not
be in conflict with the negative result of the CHOOZ long--baseline reactor 
experiment on $\bar{\nu}_e \rightarrow \bar{\nu}_\mu $ oscillations [7].

 In conclusion, our explicit model of lepton texture displays a number of 
important features. {\it (i)} It correlates correctly (with high precision) the
tauon mass with electron and muon masses. {\it (ii)} It predicts (without para%
meters) the maximal mixing between muon and tauon neutrinos in the limit 
$\mu^{(\nu)} \rightarrow 0 $, consistent with the observed deficit of atmosphe%
ric $\nu_\mu $'s. {\it (iii)} It fails to explain the observed deficit of solar
$\nu_e $'s. {\it (iv)} It predicts new $\nu_\mu \rightarrow \nu_e$ oscillations
with the amplitude consistent with LSND experiment, but with a phase corres%
ponding to the mass squared difference at least one order of magnitude smaller.

 In the framework of our model, the point {\it (iii)} may suggest that in 
Nature there exists (at least) one sort, $\nu_s^{(e)} $, of sterile neutrinos 
(blind to the Standard Model interactions), responsible for the observed 
deficit of solar $\nu_e $'s through $\nu_e \rightarrow \nu_s^{(e)} $ oscillat%
ions dominating the survival probability $P(\nu_e \rightarrow \nu_e) \simeq 1 
- P(\nu_e \rightarrow \enu )$ [8]. In an extreme version of this picture, it 
might even happen that in Nature there would be two sorts, $\nu_s^{(e)} $ and 
$\nu_s^{(\mu)} $, of sterile neutrinos, where $\nu_s^{(\mu)} $ would replace 
$\nu_\tau $ in explaining the observed deficit of atmospheric $\nu_\mu $'s by 
means of $\nu_\mu \rightarrow \nu_s^{(\mu)} $ oscillations that should dominate
the survival probability $P(\nu_\mu \rightarrow \nu_\mu) \simeq 1 - P(\nu_\mu 
\rightarrow \nu_s^{(\mu)})$ [9]. In this case, the constant $\alpha^{(\nu)}$ 
for active neutrinos might be even zero (however, very small $\alpha^{(\nu)}$ 
would be still allowed). Such a model is discussed in Sections 5 and 6.

 For the author of the present paper the idea of existence of two sorts of ste%
rile neutrinos is fairly appealing, since two such spin--1/2 fermions, blind 
to all Standard Model interactions, do follow (besides three standard families
of active leptons and quarks) [8] from the argument {\it (i)} mentioned in 
Introduction, based on the K\"{a}hler--like generalized Dirac equations. Note
in addition that the $\nu_e \rightarrow \nu_s^{(e)} $ and $\nu_\mu \rightarrow 
\nu_s^{(\mu)} $ oscillations caused by appropriate mixings should be a natural 
consequence of the spontaneous breaking of electroweak $ SU(2)\times U(1) $ 
symmetry.

 In Section 7, a possibility is considered that two extra neutrino mass 
states, whose existence is implied by two sterile neutrinos $\enu $ and $\mnu 
$, cause in the Standard Model framework some tiny neutrino instability and 
related damping of $\nu_e $ and $\nu_\mu $ oscillations.

\vspace{0.3cm}

\ni {\bf 4. Perspectives for unification with quarks}

\vspace{0.3cm}

 In this Section, we try to apply to quarks the form of mass matrix which was 
worked out above for leptons. To this end, we conjecture for three generations 
of up quarks $ u\,,\,c\,,\,t $ and down quarks $ d\,,\,s\,,\,b $ the mass 
matrices $\left(M_{\alpha \beta}^{(u)}\right)$ and $\left(M_{\alpha \beta}^{(d)}\right) $, 
respectively, essentially of the form (1), where the label $ f = u\,,\,d $ 
denotes now up and down quarks. The only modification introduced is a new real 
constant $ C^{(f)}$ added to $\varepsilon^{(f)}$ in the element $ M^{(f)}_{33}$
which now reads

\begin{equation}
M^{(f)}_{33} = \frac{24 \mu^{(f)}}{25\cdot 29}\left(624 + \varepsilon^{(f)} 
+ C^{(f)}\right)\;.
\end{equation}

 Since for quarks the mass scales $\mu^{(u)}$ and $\mu^{(d)}$ are expected to 
be even more important than the scale $\mu^{(e)}$ for charged leptons, we 
assume that the off--diagonal elements of mass matrices $\left(M_{\alpha \beta}^{(u)}
\right)$ and $\left(M_{\alpha \beta}^{(d)}\right)$ can be considered as a small 
perturbation of their diagonal terms. Then, in the lowest perturbative order, 
we obtain the following mass formulae


\begin{eqnarray}
m_{u,d} & = & \frac{\mu^{(u,d)} }{29} \varepsilon^{(u,d)} - A^{(u,d)}
\left(\frac{\alpha^{(u,d)}}{\mu^{(u,d)}}\right)^2 \; , \nonumber \\ 
m_{c,s} & = & \frac{\mu^{(u,d)}}{29} \frac{4}{9}\left(80 + \varepsilon^{(u,d)}
\right) + \left( A^{(u,d)} - B^{(u,d)} \right)\left(\frac{\alpha^{(u,d)}}{
\mu^{(u,d)} }\right)^2 \; , \nonumber \\ 
m_{t,b} & = & \frac{\mu^{(u,d)}}{29} \frac{24}{25} \left(624 + \varepsilon^{(
u,d)} + C^{(u,d)} \right) + B^{(u,d)}\left(\frac{\alpha^{(u,d)}}{\mu^{(u,d
)}}\right)^2 \;,
\end{eqnarray}


\ni where

\begin{equation}
A^{(u,d)} = \frac{\mu^{(u,d)}}{29}\,\frac{36}{320 - 5\varepsilon^{(u,d)} }\;
\;,\;\; B^{(u,d)} = \frac{\mu^{(u,d)}}{29}\,\frac{10800}{31696 + 54 C^{(u,d)}
+ 29\varepsilon^{(u,d)}}\;.
\end{equation}

\ni In Eqs. (35), the relative smallness of perturbating terms is more pronoun%
ced due to extra factors. In our discussion, we will take for experimental 
quark masses the arithmetic means of their lower and upper limits quoted in the
Review of Particle Physics [3] {\it i.e.},


\begin{equation}
m_u = 3.3 \,{\rm MeV}\;,\; m_c = 1.3 \,{\rm GeV}\;,\;m_t = 174 \,{\rm GeV}
\end{equation}

\ni and


\begin{equation}
m_d = 6 \,{\rm MeV}\;,\; m_s = 120 \,{\rm MeV}\;,\;m_b = 4.3 \,{\rm GeV}\;.
\end{equation}

 Eliminating from the unperturbed terms in Eqs. (35) the constants $\mu^{(u,d)}
$ and $\varepsilon^{(u,d)}$, we derive the correlating formulae being counter%
parts of Eqs. (2) for charged leptons:

\begin{eqnarray}
m_{t,b} & = & \frac{6}{125} \left( 351 m_{c,s} - 136 m_{u,d} \right) + 
\frac{\mu^{(u,d)}}{29}\frac{24}{25} C^{(u,d)} \nonumber \\ & & - \frac{1}{125} 
\left(2922 A^{(u,d)} - 2231 B^{(u,d)}\right) \left(\frac{\alpha^{(u,d)}}{\mu^{(
u,d)}}\right)^2\;, \nonumber \\
\mu^{(u,d)} & = & \frac{29}{320} \left(9 m_{c,s} - 4m_{u,d}\right) -
\frac{29}{320} \left(5 A^{(u,d)} - 9 B^{(u,d)}\right) \left(\frac{\alpha^{(u,d
)}}{\mu^{(u,d)}}\right)^2 \;, \nonumber \\  
\varepsilon^{(u,d)} & = & \frac{29 m_{u,d}}{\mu^{(u,d)}} + \frac{29}{\mu^{
(u,d)}} A^{(u,d)} \left(\frac{\alpha^{(u,d)}}{\mu^{(u,d)}}\right)^2 \;. 
\end{eqnarray}

\ni The unperturbed parts of these relations are:

\begin{eqnarray}
\stackrel{\circ}{m}_{t,b} & = & \frac{6}{125} \left( 351 m_{c,s} - 
136 m_{u,d} \right) + \frac{\stackrel{\circ}{\mu}^{(u,d)}}{29}\frac{24}{25} 
\stackrel{\circ}{C}^{(u,d)} \nonumber \\ & = & \left\{\begin{array}{c} 
21.9\\1.98 \end{array}\right\}\,{\rm GeV} + \frac{\stackrel{\circ}{\mu}^{(u,
d)}}{29}\frac{24}{25} \stackrel{\circ}{C}^{(u,d)} \;, \nonumber \\
\stackrel{\circ}{\mu}^{(u,d)} & = & \frac{29}{320} \left(9 m_{c,s} - 4m_{u,d}
\right) = \left\{\begin{array}{c} 1060 \\ 95.7 \end{array}\right\}\,{\rm MeV} 
\;, \nonumber \\  
\stackrel{\circ}{\varepsilon}^{(u,d)} & = & \frac{29 m_{u,d}}{\stackrel{
\circ}{\mu}^{(u,d)}} = \left\{\begin{array}{l} 0.0904 \\ 1.82 \end{array}
\right\} \;.
\end{eqnarray}

\ni In the spirit of our perturbative approach, the "coupling" constant $
\alpha^{(u,d)}$ can be put zero in all perturbing terms in Eqs. (35) and (39), 
except for $\alpha^{(u,d)\,2}$ in the numerator of the factor $(\alpha^{(u,d)}
/\mu^{(u,d)})^2$ that now becomes $(\alpha^{(u,d)}/\stackrel{\circ}{\mu}^{(u,d
)})^2$. Then, $ A^{(u,d)}$ and $ B^{(u,d)}$ are replaced by 

\begin{equation}
\stackrel{\circ}{A}^{(u,d)} = \frac{\stackrel{\circ}{\mu}^{(u,d)}}{29}
\frac{36}{320 - 5\stackrel{\circ}{\varepsilon}^{(u,d)} }\;
\;,\;\; \stackrel{\circ}{B}^{(u,d)} = \frac{\stackrel{\circ}{\mu}^{(u,d)}}{29} 
\frac{10800}{31696 + 54 \stackrel{\circ}{C}^{(u,d)} + 29\stackrel{\circ}{
\varepsilon}^{(u,d)}}\;.
\end{equation}

\ni Note that the first Eq. (35) can be rewritten identically as $m_{u,d} =
\;\stackrel{\circ}{\mu}^{(u,d)}\,\stackrel{\circ}{\varepsilon}^{(u,d)}\!\!\!/
29 $ according to the third Eq. (40).

 We shall be able to return to the discussion of quark masses after the estim%
ation of constants $\alpha^{(u)}$ and $\alpha^{(d)}$ is made. Then, we shall 
determine the parameters $ C^{(u)}$ and $C^{(d)}$ (as well as their unperturbed
parts $\stackrel{\circ}{C}^{(u)}$ and $\stackrel{\circ}{C}^{(d)}$) playing here
an essential role in providing large values for $ m_t $ and $ m_b $.

 At present, we find the unitary matrices $\left(U_{\alpha \beta}^{(u,d)}\right)$ that 
diagonalize the mass matrices $\left(M_{\alpha \beta}^{(u,d)}\right)$ according to the 
relations $ U^{(u,d)\,\dagger}M^{(u,d)}U^{(u,d)} = $ diag$(m_{u,d}\,,\,
m_{c,s}\,,\,m_{t,b})$. In the lowest perturbative order, the result has the 
form (4) with the necessary replacement of labels:


\begin{equation}
(e) \rightarrow (u)\;\;{\rm or}\;\;(d)\;,\;\mu \rightarrow c\;\;{\rm or}\;\;s
\;,\;\tau \rightarrow t\;\;{\rm or}\;\;b\;,
\end{equation}

\ni respectively.

 Then, the elements $ V_{\alpha \beta}$ of the \CKM matrix $ V = U^{(u)\,\dagger}U^{(d)}$ 
can be calculated with the use of Eqs. (42) in the lowest perturbative order. 
Six resulting off--diagonal elements are:


\begin{eqnarray}
V_{us} & = & -V^*_{cd} = \frac{2}{29}\left(\frac{\alpha^{(d)}}{m_s} 
e^{i\varphi^{(d)}} - \frac{\alpha^{(u)}}{m_c} e^{i\varphi^{(u)}} \right) \;,
\nonumber \\ 
V_{cb} & = & -V^*_{ts} = \frac{8\sqrt{3}}{29}\left(\frac{\alpha^{(d)}}{m_b} 
e^{i\varphi^{(d)}} - \frac{\alpha^{(u)}}{m_t} e^{i\varphi^{(u)}} \right) \simeq
\frac{8\sqrt{3}}{29} \frac{\alpha^{(d)}}{m_b} e^{i\varphi^{(d)}} \;, 
\nonumber \\
V_{ub} & \simeq & -\frac{16\sqrt{3}}{841}\frac{\alpha^{(u)}\alpha^{(d)} }{
m_c m_b} e^{i(\varphi^{(u)}+\varphi^{(d)})} \;, \nonumber \\ 
V_{td} & \simeq & \frac{16\sqrt{3}}{841} 
\frac{\alpha^{(d)\,2}}{m_s m_b}\,e^{-2i\varphi^{(d)}} \;,
\end{eqnarray}

\ni where the indicated approximate steps were made due to the inequality $ m_t
\gg m_b $ and/or under the assumption that $\alpha^{(u)}/m_c \gg \alpha^{(d)}
/m_b $ [{\it cf.} the conjecture (46)]. All three diagonal elements are real 
and positive in a good approximation:

\begin{equation}
V_{ud} \simeq 1 - \frac{1}{2}|V_{us}|^2\;,\;V_{cs} \simeq 1 - \frac{1}{2}
|V_{us}|^2 - \frac{1}{2}|V_{cb}|^2\;,\;V_{tb} \simeq 1 - \frac{1}{2}|V_{cb}|^2
\;.
\end{equation}

\ni In fact, in the lowest perturbative order,


\begin{equation}
\arg V_{ud} \simeq \frac{4}{841} \frac{\alpha^{(u)}\alpha^{(d)}}{m_c m_s}
\sin \left(\varphi^{(u)} - \varphi^{(d)}\right)\frac{180^\circ}{\pi} \simeq 
-\arg V_{cs}\;,\;\arg V_{tb} \simeq 0 \;,
\end{equation}

\ni what gives $ \arg V_{ud} = 0.88^\circ = -\arg V_{cs}$, if the values (46), 
(49) and (52) are used.

 Taking as an input the experimental value $|V_{cb}| = 0.0395 \pm 0.0017 $ [3],
we estimate from the second Eq. (43) that 

\begin{equation}
\alpha^{(d)} \simeq \frac{29}{8\sqrt{3}}\, m_b\, |V_{cb}| = (355 \pm 15)\;{\rm 
MeV} \;,
\end{equation}

\ni where $ m_b = 4.3 $ GeV. In order to estimate also $\alpha^{(u)}$, we will
tentatively conjecture the approximate proportion

\begin{equation}
\alpha^{(u)} : \alpha^{(d)} \simeq Q^{(u)\,2} : Q^{(d)\,2} = 4
\end{equation}

\ni to hold, where $ Q^{(u)} = 2/3 $ and $ Q^{(d)} = -1/3 $ are quark electric 
charges. Note that in the case of leptons we had $\alpha^{(\nu)} : \alpha^{(e)}
= 0.37 : (\sqrt{180}\,\times 10^6) = 2.8\times 10^{-8}$ for the central value of
$\alpha^{(e)}$ [{\it cf.} Eqs. (3) and (30)], what is consistent with the 
analogical approximate proportion

\begin{equation}
\alpha^{(\nu)} : \alpha^{(e)} \simeq Q^{(\nu)\,2} : Q^{(e)\,2} = 0 \;,
\end{equation}

\ni where $Q^{(\nu)} = 0 $ and $ Q^{(e)} = -1 $ are lepton electric charges. 
Under the conjecture (47):

\begin{equation}
\alpha^{(u)} \simeq (1420 \pm 60)\, {\rm MeV} \;.
\end{equation}

\ni In this case, from the second and third Eq. (43) we obtain the prediction

\begin{equation}
|V_{ub}|/|V_{cb}| \simeq \frac{2}{29}\frac{\alpha^{(u)}}{m_c} \simeq 0.0753 
\pm 0.0032 \;,
\end{equation}

\ni where $ m_c = 1.3 $ GeV. This is consistent with the experimental 
figure $|V_{ub}|/|V_{cb}| = 0.08 \pm 0.02 $ [3].

 Now, with the experimental value $|V_{us}| = 0.2196 \pm 0.0023$ [3] as 
another input, we can calculate from the first Eq. (43) the phase difference 
$\varphi^{(u)} - \varphi^{(d)}$. In fact, taking the absolute value of this 
equation, we get

\begin{equation}
\cos\left(\varphi^{(u)} - \varphi^{(d)}\right) = \frac{1}{8}\frac{m_c}{m_s}
\left[1 + 16\left(\frac{m_s}{m_c}\right)^2 - \frac{841}{4}\left(\frac{m_c}{
\alpha^{(d)}}\right)^2 |V_{us}|^2 \right] = - 0.0301 
\end{equation}

\ni with $ m_c = 1.3 $ GeV  and $ m_s = 120 $ MeV, if the proportion (47) is 
taken into account. Here, the central values of $\alpha^{(d)}$ and $|V_{us}|$ 
were used. Hence,

\begin{equation}
\varphi^{(u)} - \varphi^{(d)} = 91.7^\circ = -88.3^\circ + 180^\circ
\end{equation}

\ni so, this phase difference turns out to be near $ 90^\circ $. Then, 
calculating the argument of the first Eq. (43), we infer that

\begin{equation}
\tan\left(\arg V_{us} - \varphi^{(d)}\right) = -4 \,\frac{m_s}{m_c}\,
\frac{\sin\left(\varphi^{(u)} - \varphi^{(d)}\right)}{1 - 4 ({m_s}/{m_c}) 
\cos\left(\varphi^{(u)} - \varphi^{(d)}\right)} = - 0.365 \;,
\end{equation}

\vspace{0.1cm}

\ni what gives


\begin{equation}
\arg V_{us} = -20.1^\circ + \varphi^{(d)} \;.
\end{equation}

 The results (52) and (54) together with the formula (43) enable us to evaluate
the rephasing--invariant CP--violating phases

\begin{equation}
\arg (V_{us}^*V_{cb}^*V_{ub}) = 20.1^\circ - 88.3^\circ = -68.2^\circ 
\end{equation}

\vspace{0.1cm}

\ni and 


\begin{equation}
\arg (V_{cd}^*V_{ts}^*V_{td}) = -20.1^\circ \;,
\end{equation}

\ni which turn out to be near to -70$^\circ$ and -20$^\circ$, respectively
(they are invariant under quark rephasing equal for up and down quarks of the 
same generation). Note that the sum of arguments (55) and (56) is always equal 
to $\varphi^{(u)} - \varphi^{(d)} - 180^\circ $. Carrying out quark rephasing 
(equal for up and down quarks of the same generation), where

\begin{equation}
\arg V_{us} \rightarrow 0 \;,\; \arg V_{cb} \rightarrow 0 \;,\; \arg V_{cd}
\rightarrow 180^\circ \;,\; \arg V_{ts} \rightarrow 180^\circ
\end{equation}

\ni and $\arg V_{ud}$, $\arg V_{cs}$, $\arg V_{tb}$ remain unchanged, we
conclude from Eqs. (55) and (56) that

\begin{equation}
\arg V_{ub} \rightarrow -68.2^\circ \;,\; \arg V_{td} \rightarrow -20.1^\circ
\;.
\end{equation}

\ni The sum of arguments (58) after rephasing (57) is always equal to $
\varphi^{(u)} - \varphi^{(d)} - 180^\circ $.

 Thus, in this quark phasing, we predict the following \CKM matrix:

\begin{equation}
\left( V_{\alpha \beta} \right) = \left(\begin{array}{ccc}
0.976 & 0.220 & 0.00297\,e^{-i\,68.2^\circ}\\ -0.220 & 0.975 & 0.0395 \\  
0.00805\,e^{-i\,20.1^\circ} & -0.0395 & 0.999 \end{array}\right)\;.
\end{equation}

\ni Here, only $|V_{us}|$ and $|V_{cb}|$ [and quark masses $m_s\,,\;m_c\,,\;
m_b $ consistent with the mass matrices $\left( M^{(u)}_{\alpha \beta}\right)$ 
and $\left( M^{(d)}_{\alpha \beta}\right)$] are our inputs, while all other 
matrix elements $ V_{\alpha \beta}$, partly induced by unitarity, are evaluated
from the relations derived in this Section from the Hermitian mass matrices $
\left( M^{(u)}_{\alpha \beta}\right)$ and $\left( M^{(d)}_{\alpha \beta}\right)
$ [and the conjectured proportion (47)]. The independent predictions are $|V_{
ub}|$ and arg$ V_{ub}$. In Eq. (59), the small phases arising from Eqs. (45), 
$\arg V_{ud} = 0.9^\circ$ and $\arg V_{cs} = -0.9^\circ$, are neglected (here, 
arg $(V_{ud}V_{cs}V_{tb}) = 0 $).

 The above prediction of $ V_{\alpha \beta}$ implies the following values of 
Wolfenstein parameters [3]:

\begin{equation}
\lambda = 0.2196\;\;,\;\;A = 0.819 \;\;,\;\;\rho = 0.127\;\;,\;\;\eta = 0.319 
\end{equation}

\ni and of unitary--triangle angles:

\begin{equation}
\gamma = \arctan \frac{\eta}{\rho} = - \arg V_{ub} = 68.2^\circ\;\;,\;\;\beta
= \arctan \frac{\eta}{1-\rho} = - \arg V_{td} = 20.1^\circ \;.
\end{equation}

\ni The predicted large value of $\gamma $ follows the present experimental 
tendency.

 If instead of the central value $|V_{us}| = 0.2196$ we take as the input 
the range $|V_{us}| = 0.2173$ to 0.2219, we obtain from Eq. (51) $\varphi^{(u)}
- \varphi^{(d)} = 89.8^\circ\;\;{\rm to}\;\;93.6^\circ$ (with $|V_{cb}| = 
0.0395$ giving $\alpha^{(d)} = 355$ MeV), what implies through Eq. (53) that 
arg $V_{us}- \varphi^{(d)} = {-20.3}^\circ\;\;{\rm to}\;\; {-19.8}^\circ $. 
Then, after rephasing (57), arg$V_{ub} = -69.9^\circ\;\;{\rm to}\;\;-66.6^\circ
$ and ${\arg V_{td}} = -20.3^\circ\;\;{\rm to}\;\;{-19.8}^\circ$. In this case,
the Wolfenstein parameters are $\lambda = 0.2173 $ to 0.2219, $ A = 0.837 $ to 
0.802, $\rho = 0.119 $ to 0.135 and $\eta = 0.325 $ to 0.312 (here, $\lambda
\sqrt{\rho^2 + \eta^2} = |V_{ub}|/|V_{cb}| = 0.0753 $ is fixed). Thus, $\gamma 
= - \arg V_{ub} = 69.9^\circ\;\;{\rm to}\;\;66.6^\circ$ and $\beta = - \arg 
V_{td} = 20.3^\circ \;\;{\rm to}\;\;19.8^\circ$.

 In contrast, if the central value $|V_{cd}| = 0.0395$ (giving $\alpha^{(d)} = 
355$ MeV) is replaced by the input of the range $V_{cd} = 0.0378$ to 0.0412 
(corresponding to $\alpha^{(d)} = 340$ to 370 MeV), we calculate from Eq. (51) 
that $\varphi^{(u)} - \varphi^{(d)} = 97.3^\circ\;\;{\rm to}\;\;84.9^\circ$ 
(with $|V_{us}| = 0.2196 $), what leads to arg$V_{us} - \varphi^{(d)} = -
19.3^\circ\;\;{\rm to}\;\;-20.9^\circ$. Hence, after rephasing (57), arg$V_{ub}
= -63.4^\circ\;\;{\rm to}\;\;{-74.6^\circ}$ and ${\arg V_{td}} = -19.3^\circ\;\;
{\rm to}\;\;-20.9^\circ$. In this case, the Wolfenstein parameters take the 
values $\lambda = 0.2196 $, $A = 0.784 $ to 0.854, $\rho = 0.149$ to 0.0951 and
$\eta = 0.298 $ to 0.345. Thus, $\gamma = - \arg V_{ub} = 63.4^\circ\;\;{\rm 
to} \;\;74.6^\circ$ and $\beta = - \arg V_{td} = 19.3^\circ \;\;{\rm to}\;\;
20.9^\circ$. Here, $|V_{ub}| = 0.00273 $ to 0.00323 and $|V_{td}| = 0.00738 $ 
to 0.00874.

  Eventually, we may turn back to quark masses. From the third Eq. (35) we can 
evaluate

\begin{equation}
C^{(u,d)} = \frac{29}{\mu^{(u,d)}}\,\frac{25}{24}\,m_{t,b} - 624 - 
\varepsilon^{(u,d)} - \frac{29}{\mu^{(u,d)}}\,\frac{25}{24}\,B^{(u,d)}\left(\frac
{\alpha^{(u,d)}}{\mu^{(u,d)}}\right)^2\;,
\end{equation}

\ni what, in the framework of our perturbative approach, gives

\begin{eqnarray}
C^{(u,d)} & = & \stackrel{\circ}{C}^{(u,d)} + \frac{29}{\stackrel{\circ}{\mu}^{
(u,d)}}\, \frac{25}{24}\,m_{t,b} \,\frac{29}{320\stackrel{\circ}{\mu}^{(u,d)}}
\,\left(5\stackrel{\circ}{A}^{(u,d)} - 9\stackrel{\circ}{B}^{(u,d)}\right)\,
\left(\frac{\alpha^{(u,d)}}{\stackrel{\circ}{\mu}^{(u,d)}}\right)^2 \nonumber
\\ & & - \frac{29}{\stackrel{\circ}{\mu}^{(u,d)}}\,\left(
\stackrel{\circ}{A}^{(u,d)} + \stackrel{\circ}{B}^{(u,d)}\right)\,\left(
\frac{\alpha^{(u,d)}}{\stackrel{\circ}{\mu}^{(u,d)}}\right)^2\;,
\end{eqnarray}

\ni where

\begin{equation}
\stackrel{\circ}{C}^{(u,d)} = \frac{29}{\stackrel{\circ}{\mu}^{(u,d)}}\, 
\frac{25}{24}\,m_{t,b} - 624 - \stackrel{\circ}{\varepsilon}^{(u,d)} =
\left\{\begin{array}{c} 4339 \\ 733.2 \end{array}\right\} =
\left\{\begin{array}{r} 4340 \\ 733 \end{array}\right\} \;.
\end{equation}

\ni With the central values of $\alpha^{(u)}$ and $\alpha^{(d)}$ as estimated 
in Eqs. (46) and (49) we find from Eqs. (41)

\begin{equation}
\stackrel{\circ}{A}^{(u,d)}\left(\frac{\alpha^{(u,d)}}{\stackrel{\circ}{\mu
}^{(u,d)}}\right)^2 = \left\{\begin{array}{r} 7.39 \\ 5.26 \end{array}
\right\}\,{\rm MeV}\;,\;
\stackrel{\circ}{B}^{(u,d)}\left(\frac{\alpha^{(u,d)}}{\stackrel{\circ}{\mu
}^{(u,d)}}\right)^2 = \left\{\begin{array}{r} 2.66 \\ 6.88 \end{array}
\right\}\,{\rm MeV} \;,
\end{equation}

\ni where

\begin{equation}
\frac{\stackrel{\circ}{\mu}^{(u,d)}}{29}\left(\frac{\alpha^{(u,d)}}{\stackrel{
\circ}{\mu}^{(u,d)}}\right)^2 = \left\{\begin{array}{r} 65.6 \\ 45.4 
\end{array}\right\}\,{\rm MeV}\;.
\end{equation}

\ni We calculate from Eqs. (63) with the use of values (65) that

\begin{equation}
C^{(u,d)} = \left\{\begin{array}{c} 4339 + 5.25 \\ 733.2 - 49.5 \end{array}
\right\} = \left\{\begin{array}{c} 4344 \\ 683.7 \end{array}
\right\}  = \left\{\begin{array}{r} 4340 \\ 684 \end{array}
\right\} \;.
\end{equation}

 Similarly, from the second Eq. (39), making use of the values (65), we obtain

\begin{equation}
{\mu}^{(u,d)} = \left\{\begin{array}{c} 1060 - 1.18 \\ 95.7 + 3.23 \end{array}
\right\}\,{\rm MeV} = \left\{\begin{array}{c} 1059 \\ 98.9 \end{array}\right\}
\,{\rm MeV} = \left\{\begin{array}{c} 1060 \\ 98.9 \end{array}\right\}
\,{\rm MeV}\;.
\end{equation}

 We can easily check that, with the values (40) for $\stackrel{\circ}{\mu}^{(u,
d)}$ and $\stackrel{\circ}{\varepsilon}^{(u,d)}$ and the value (64) for $
\stackrel{\circ}{C}^{(u,d)}$ determined as above from quark masses, the 
unperturbed parts of mass formulae (35) reproduce correctly these masses. In 
fact,

\begin{eqnarray}
\stackrel{\circ}{m}_{u,d} & = &  \frac{\stackrel{\circ}{\mu}^{(u,d)}}{29}\,
\stackrel{\circ}{\varepsilon}^{(u,d)} = \left\{\begin{array}{c} 3.3 \\ 6 
\end{array}\right\}\,{\rm MeV}\;, \nonumber \\
\stackrel{\circ}{m}_{c,s} & = &  \frac{\stackrel{\circ}{\mu}^{(u,d)}}{29}
\,\frac{4}{9}\left(80 + \stackrel{\circ}{\varepsilon}^{(u,d)}\right) =
\left\{\begin{array}{r} 1300 \\ 120 \end{array}\right\}\,{\rm MeV}\;, 
\nonumber \\
\stackrel{\circ}{m}_{t,b} & = &  \frac{\stackrel{\circ}{\mu}^{(u,d)}}{29}
\,\frac{24}{25}\left(624 + \stackrel{\circ}{\varepsilon}^{(u,d)} + 
\stackrel{\circ}{C}^{(u,d)}\right) = \left\{\begin{array}{c} 174 \\ 4.3 
\end{array}\right\}\,{\rm GeV}\;.
\end{eqnarray}

\ni The same is true for the unperturbed part of the first correlating formula
(39). The --- here omitted --- corrections to Eqs. (69), arising from all pert%
urbing terms in the mass formulae (35) (including the corrections from ${\delta
\mu^{(u,d)}} $, $\delta \varepsilon^{(u,d)}$ and $\delta C^{(u,d)}$), are 
relatively small, {\it viz.}

\begin{equation}
\delta m_{u,d}\! = \!\left\{\begin{array}{r}3.7\times 10^{-3} \\ -2.0\times 
10^{-1}\end{array}\right\}\,{\rm MeV}\,,\,\delta m_{c,s}\! = \!\left\{
\begin{array}{r}9.5 \\ -3.8 \end{array}\right\}\,{\rm MeV}\,,\,\delta m_{t,b}\!
= \!\left\{\begin{array}{r} 170 \\ -74 \end{array}\right\}\,{\rm MeV}\,, 
\end{equation}

\ni respectively.

We would like to stress that, in contrast to the case of charged leptons, where
$m_\tau $ has been predicted from $m_e $ and $m_\mu $, in the case of up and 
down quarks two extra parameters $ C^{(u)}$ and $ C^{(d)}$ appear necessarily 
to provide large masses $m_t $ and $m_b $ (much larger than $m_\tau $). They 
cause that $m_t $ ($m_b $) cannot be predicted from $m_u $ and $m_c $ ($m_d $ 
and $m_s $), till the new parameters are quantitatively understood.

 Note that a conjecture about $ C^{(u)}$ and $ C^{(d)}$ might lead to a predic%
tion for quark masses and so, introduce changes in the "experimental" quark 
masses (37) and (38) accepted here. The same is true for a conjecture about $ 
\varphi^{(u)}$ and $\varphi^{(d)}$.

 For instance, the conjecture that the phase difference $\varphi^{(u)} - 
\varphi^{(d)}$ is maximal,

\begin{equation}
\varphi^{(u)} - \varphi^{(d)} = 90^\circ \;,
\end{equation}

\vspace{0.1cm}

\ni leads through the first equality in Eq. (51) to the condition

\begin{equation}
1 + 16\left(\frac{m_s}{m_c}\right)^2 - \frac{841}{4}\left(\frac{m_s}{
\alpha^{(d)}}\right)^2 |V_{us}|^2 = 0
\end{equation}

\vspace{0.1cm}

\ni predicting for $ s $ quark the mass

\begin{equation}
m_s = 118.7\,{\rm MeV} = 119\,{\rm MeV} 
\end{equation}

\vspace{0.1cm}

\ni (with $\alpha^{(d)} = 355 $ MeV), being only slightly lower than the value 
120 MeV used previously. Here, $ m_c $ and $ m_b $ are kept equal to 1.3 and 4.3
GeV, respectively (also masses of $ u\,,\,d $ and $t $ quarks are not changed, 
while $\stackrel{\circ}{\mu}^{(d)}$, $\stackrel{\circ}{\varepsilon}^{(d)}$ and 
$\stackrel{\circ}{C}^{(d)}$ change slightly). Then, from the first equality in 
Eq. (53)

\begin{equation}
\tan\left(\arg V_{us} - \varphi^{(d)}\right) = -4 \,\frac{m_s}{m_c}
= - 0.365 \;\;,\;\;\arg V_{us} = -20.1^\circ + \varphi^{(d)} \;.
\end{equation}

\vspace{0.1cm}

\ni After rephasing (57), this gives $\arg V_{ub} + \arg V_{td} = \varphi^{(u)}
- \varphi^{(d)} - 180^\circ = -90^\circ $, where

\begin{equation}
\arg V_{ub} = -69.9^\circ\;\;,\;\;\arg V_{td} = -20.1^\circ
\end{equation}

\vspace{0.1cm}

\ni {\it i.e.}, practically $-70^\circ$ and $-20^\circ$. All $ |V_{\alpha \beta
}|$ remain unchanged (with our inputs of $ |V_{us}| = 0.2196 $ and $ |V_{cb}| 
= 0.0395 $), except for $ |V_{td}| $ which changes slightly, becoming

\begin{equation}
|V_{td}|  = 0.00814\;.
\end{equation}

\vspace{0.1cm}

\ni Thus, in the \CKM matrix predicted in Eq. (59), only $ |V_{td}| $ and the
phases (75) show some changes. The Wolfenstein parameters are

\begin{equation}
\rho = 0.118 \;\;,\;\; \eta = 0.322
\end{equation}

\vspace{0.1cm}

\ni and $\lambda $ and $ A $ unchanged (here, the sum $\rho^2 + \eta^2 = 0.118$
is also unchanged). Hence, $\gamma + \beta = 90^\circ $ and $\alpha = 180^\circ
- \gamma - \beta = 90^\circ $, where

\begin{equation}
\gamma = \arctan \frac{\eta}{\rho} = - \arg V_{ub} = 69.9^\circ \;\;,\;\;
\beta = \arctan \frac{\eta}{1 - \rho} = - \arg V_{td} = 20.1^\circ\;.
\end{equation}

\ni So, in the case of conjecture (71), the new restrictive relation

\begin{equation}
\frac{\eta}{\rho} = \frac{1 - \rho}{\eta}\;\;,\;\; \rho^2 + \eta^2 = \rho
\end{equation}

\ni holds, implying the prediction

\begin{equation}
|V_{td}| /|V_{ub}| = \sqrt{\frac{(1-\rho)^2 + \eta^2}{\rho^2 + \eta^2}} =
\frac{\eta}{\rho} = 2.74 \;,
\end{equation}

\ni due to the definition of $\rho $ and $\eta $ from $ V_{ub}$ and $ V_{td}$.
It is in agreement with our figures for $ |V_{td}|$ and $ |V_{ub}|$. Then, the 
new relationship

\begin{equation}
\frac{1}{4}\frac{m_c}{m_s}= \frac{\alpha^{(d)} m_c}{\alpha^{(u)} m_s} = 
\frac{\eta}{\rho} 
\end{equation}

\ni follows for quark masses $m_c $, $m_s $ and Wolfenstein parameters $\rho $,
$\eta $, in consequence of Eqs. (43) and the conjectured proportion (47). 
Both its sides are really equal for our values of $m_c $, $m_s$ and $\rho $,
$\eta $.

 Thus, summarizing, we cannot predict quark masses without an {\it additional}
knowledge or conjecture about the constants $\mu^{(u,d)}$, $\varepsilon^{(u,d)
}$, $ C^{(u,d)}$, $\alpha^{(u,d)}$ and $\varphi^{(u,d)}$ (in particular, the
conjecture (71) predicting $ m_s $ may be natural). However, we always describe
them correctly. If we describe them {\it jointly} with quark mixing parameters,
we obtain two independent predictions of $|V_{ub}|$ and $\gamma = - \arg V_{ub}
$: the whole \CKM matrix is calculated from the inputs of $|V_{us}|$ and of $|
V_{ub}| $ [and of quark masses $ m_s $, $ m_c $ and $ m_b $ consistent with 
the mass matrices $\left( M^{(u)}_{\alpha \beta}\right) $ and $\left( M^{(d)}_{
\alpha \beta}\right) $].

 Concluding this Section, we can claim that our leptonic form of mass matrix 
works also in a promising way for up and down quarks. But, it turns out that, 
in the framework of the leptonic form of mass matrix, the heaviest quarks, $ t 
$ and $ b $, require an additional mechanism in order to produce the bulk of 
their masses (here, it is represented by the large constants $ C^{(u)}$ and 
$ C^{(d)}$). Such a mechanism, however, intervenes into the process of quark 
mixing only through quark masses (practically $m_t$ and $m_b$) and so, it does 
not modify for quarks the leptonic form of mixing mechanism.

\vspace{0.35cm}

\ni {\bf 5. A model of texture with two sterile neutrinos}

\vspace{0.35cm}

 Assume that there are two sorts, $\enu $ and $\mnu $, of sterile neutrinos 
(blind to all Standard Model interactions and so, interacting only gravitation%
ally). Conjecture that their mixings with two active neutrinos $\nu_e $ and 
$\nu_\mu $, respectively, dominate all neutrino mixings. Thus, five flavor 
neutrino fields, $\nu_\alpha = \nu_e $, $\nu_\mu $, $\nu_\tau $, $\enu $, $\mnu
$, exist in this texture and mix according to a neutrino mass matrix $ 
M^{(\nu)}$. This can be assumed consistently in the following $ 5\times 5 $ 
form:


\begin{equation}
M^{(\nu)} =\left(M^{(\nu)}_{\alpha \beta}\right) = \left( \begin{array}{ccccc}
M_{11}^{(\nu)} & 0 & 0 & M_{14}^{(\nu)} & 0 \\ 0 & M_{22}^{(\nu)} & 0 & 0 & 
M_{25}^{(\nu)} \\ 0 & 0 & M_{33}^{(\nu)} & 0 & 0 \\ M_{41}^{(\nu)} & 0 & 0 & 0 
& 0 \\ 0 & M_{52}^{(\nu)} & 0 & 0 & 0 \end{array} \right) 
\end{equation}

\vspace{0.1cm}

\ni with $ M_{\alpha \beta}^{(\nu)} = M_{\beta \alpha}^{(\nu)*}$, $ M_{\alpha 
\alpha}^{(\nu)} = |M_{\alpha \alpha}^{(\nu)}|$ and $M_{\alpha \beta}^{(\nu)} 
= |M_{\alpha \beta}^{(\nu)}| \exp \left(i\varphi^{(\nu)}\right)$ for $ \alpha 
< \beta $, where the diagonal elements $M_{11}^{(\nu)}$, $ M_{22}^{(\nu)}$ 
and $ M_{33}^{(\nu)}$ are given in terms of $\mu^{(\nu)}$ and $\varepsilon^{(
\nu)}$ as in Eq. (1) (with $ f = \nu $). Here, we put $ M_{44}^{(\nu)} = 0 = 
M_{55}^{(\nu)}$ and even $ M_{12}^{(\nu)} = 0 = M_{23}^{(\nu)}$, the latter 
implying $\alpha^{(\nu)} = 0 $ due to Eq. (1) (with $ f = \nu $). With such a 
specific ansatz as (82), all neutrino mixings are caused by the existence of 
sterile neutrinos responsible for the off-diagonal matrix elements $ 
M_{14}^{(\nu)} $ and $ M_{25}^{(\nu)}$.

 It is important to notice that, according to the useful formula for electric
charge, $ Q = I^L_3 + Y/2 $ with $ Y/2 = I^R_3 +(B - L)/2 $, sterile neutrinos
can carry no lepton number, $ L = 0 $. This may be a reason for $ M_{44}^{(\nu
)} = 0 = M_{55}^{(\nu)}$. On the other hand, the off-diagonal matrix elements 
$ M_{14}^{(\nu)}$ and $ M_{25}^{(\nu)}$, if nonzero, violate the lepton number 
conservation.

 The mass matrix of the form (82) leads to the following masses corresponding 
to five mass neutrino fields $\nu_i = \nu_1\,,\,\nu_2\,,\,\nu_3\,,\,\nu_4\,,\,
\nu_5 $: 


\begin{eqnarray}
m_{\nu_1,\,\nu_4} & = & \frac{M^{(\nu)}_{11}}{2} \pm \sqrt{\left(\frac{M^{(\nu
)}_{11}}{2}\right)^2 + |M^{(\nu)}_{14}|^2}
\;, \nonumber \\ m_{\nu_3\,}\;\;\;\, & = & M^{(\nu)}_{33}\;, \nonumber \\
m_{\nu_2,\,\nu_5} & = & \frac{M^{(\nu)}_{22}}{2} \pm \sqrt{
\left(\frac{M^{(\nu)}_{22}}{2}\right)^2 + |M^{(\nu)}_{25}|^2}
\;.
\end{eqnarray}

\vspace{-0.1cm}

\ni Note that in Eq. (82) we used for simplicity $\alpha = 1\,,\,2\,,\,3\,,
\,4\,,\,5 $, which convention, if used properly, does not introduce any serious
confusion with $ i = 1\,,\,2\,,\,3\,,\,4\,,\,5 $.

 The corresponding $5\times 5$ unitary matrix $ U^{(\nu)}$, diagonalizing the 
neutrino mass matrix (82) according  to the relation $\,U^{(\nu)\,\dagger}\, 
M^{(\nu)}\,U^{(\nu)} = {\rm diag}(m_{\nu_1}\,,\,m_{\nu_2}\,,\,m_{\nu_3}\,,\,
m_{\nu_4}\,,\,m_{\nu_5})$~, takes the form

\vspace{-0.1cm}

\begin{equation}
U^{(\nu)} = \left({U}^{(\nu)}_{\alpha i}\right) = \left(\begin{array}{ccccc}
\frac{1}{\sqrt{1+X^2}} & 0 & 0 & -\frac{X}{\sqrt{1+X^2}} e^{i \varphi^{(\nu)}} 
& 0  \\ 0 & \frac{1}{\sqrt{1+Y^2}} & 0 & 0 & -\frac{Y}{\sqrt{1+Y^2}} e^{i
\varphi^{(\nu)}} \\  0 & 0 & 1 & 0 & 0  \\ \frac{X}{\sqrt{1+X^2}} e^{-i
\varphi^{(\nu)}} & 0 & 0 & \frac{1}{\sqrt{1+X^2}} & 0 \\ 0 & \frac{Y}{\sqrt{
1+Y^2}} e^{-i \varphi^{(\nu)}} & 0 & 0 & \frac{1}{\sqrt{1+Y^2}} 
\end{array}\right) \,,
\end{equation}

\vspace{-0.1cm}

\ni where 

\vspace{-0.1cm}

\begin{eqnarray}
X & = & \frac{m_{\nu_1} - M_{11}^{(\nu)}}{|M_{14}^{(\nu)}|} = -\frac{M_{11}^{(
\nu)}}{2|M_{14}^{(\nu)}|} + \sqrt{1 + \left(\frac{M_{11}^{(\nu)}}{2|M_{14}^{(
\nu)}|}\right)^2}\;, \nonumber \\ & & \nonumber \\ Y & = & \frac{m_{\nu_5} - 
M_{22}^{(\nu)}}{|M_{25}^{(\nu)}|} = - \frac{M_{22}^{(\nu)}}{2|M_{25}^{(\nu)}|}
+ \sqrt{1 + \left(\frac{M_{22}^{(\nu)}}{2|M_{25}^{(\nu)}|} \right)^2} \;.
\end{eqnarray}

\vspace{-0.1cm}

\ni Note that always $ 0< X \leq 1 $ and $ 0 < Y \leq 1 $.

 The flavor neutrino fields $\nu_\alpha $ are connected to the mass neutrino 
fields $\nu_i $ through the five--dimensional unitary transformation

\vspace{-0.1cm}

\begin{equation}
\nu_\alpha = \sum_i (V^\dagger)_{\alpha i}\,\nu_i
\end{equation}

\vspace{-0.1cm}

\ni with $\left(V^\dagger\right)_{\alpha i} = \left( V\right)_{i \alpha}^{*}
= V^{^*}_{i \alpha} $, where $ V = \left(V_{i \alpha}\right)$ denotes the 
lepton $ 5\times 5 $ counterpart of \CKM matrix:

\vspace{-0.1cm}

\begin{equation}
V = U^{(\nu)\,\dagger}U^{(e)}\;\;,\;\; U^{(e)} = \left( U^{(e)}_{\alpha \beta}
\right) = \left(\begin{array}{cc} U^{(e)}_{\alpha \beta}\;\,(\alpha,\beta =
1,2,3) & 0 \\ & \\ 0 & \delta_{\alpha \beta}\;\,(\alpha,\beta = 4,5) 
\end{array}\right)\;,
\end{equation}

\vspace{-0.1cm}

\ni where $\left(U^{(e)}_{\alpha \beta}\;\,(\alpha,\,\beta = 1,2,3)\right)$ is 
the charged--lepton diagonalizing unitary matrix given perturbatively in Eq. 
(4). If there $\alpha^{(e)}/\mu^{(e)}$ (jointly with its numerical coef\-%
ficients) is neglected, then $ U^{(e)} \simeq \left(\delta_{\alpha \beta}
\right)$ and so, we can put in Eq. (86)

\vspace{0.1cm}

\begin{equation}
V_{i \alpha}^* = \left(V^{\dagger}\right)_{\alpha i} = 
\left( U^{(e)\,\dagger} U^{(\nu)} \right)_{\alpha i} \simeq \left( U^{(\nu)}
\right)_{\alpha i} = U^{(\nu)}_{\alpha i}\;.
\end{equation}

\vspace{0.1cm}

\ni In our model, $ U^{(\nu)}_{\alpha i}$ are given as in Eq. (84).

\vspace{0.35cm}

\ni {\bf 6. Neutrino oscillations and their possible damping}

\vspace{0.35cm}

 Having once found the extended \CKM matrix $ V $, we can calculate the 
probabilities $ P(\nu_\alpha \rightarrow \nu_\beta)$ of neutrino oscillations 
$\nu_\alpha \rightarrow \nu_\beta $ (in the vacuum) {\it i.e.}, the probabil%
ities of (vacuum) oscillations of the flavor neutrino states $|\nu_\alpha
\rangle \rightarrow |\nu_\beta \rangle $, where $|\nu_\alpha \rangle = 
\nu_\alpha^\dagger|0 \rangle $ and

\vspace{0.1cm}

\begin{equation}
|\nu_\alpha \rangle  = \sum_i |\nu_i \rangle V_{i \alpha } 
\end{equation}

\vspace{0.1cm}

\ni with $|\nu_i \rangle = \nu_i^\dagger|0 \rangle $. If allowing that, in 
general, not all mass neutrino states $|\nu_i \rangle$ are {\it absolutely} 
stable, then

\vspace{0.1cm}

\begin{equation}
|\nu_i (t) \rangle  = e^{-i(H -i\Gamma)t}|\nu_i \rangle = |\nu_i \rangle
e^{-i(E_i -i\gamma_i)t}\;,
\end{equation}

\vspace{0.1cm}

\ni where $ E_i = \sqrt{\vec{p}^{\,2} + m^2_{\nu_i}} \simeq |\vec{p}| + m^2_{
\nu_i}/2|\vec{p}| $ and $\gamma_i = \left(|m_{\nu_i}|/E \right)\gamma^{(0)}_i
$ are neutrino energies and decay widths (with $\gamma_i^{(0)} $ and $ E \simeq
|\vec{p}|$ denoting the neutrino decay widths at rest and neutrino beam energy,
respectively). Thus, generally, we obtain for neutrinos (in the vacuum) the 
following {\it damped} oscillation formulae:


\begin{eqnarray}
P(\nu_\alpha \rightarrow \nu_\beta) & = & |\langle \nu_\beta|
e^{-i(H -i\Gamma)t}|\nu_\alpha \rangle |^2 = \sum_{j\,i}V_{j \beta}V_{j 
\alpha}^* V_{i \beta}^* V_{i \alpha} e^{i(E_j - E_i)t} e^{-(\gamma_j +
\gamma_i)t} \nonumber \\ & = & \delta_{\beta \alpha} + \sum_{j\,i} V_{j 
\beta}V_{j \alpha}^* V_{i \beta}^* V_{i \alpha} \left[ e^{i(E_j - E_i)t} 
e^{-(\gamma_j + \gamma_i)t} - 1 \right]\;.
\end{eqnarray}


\ni They are analogues of the formulae for $ K^0 \rightarrow \overline{K}^0 $ 
and $\overline{K}^0 \rightarrow K^0 $ oscillations. Note that Eqs. (91) 
imply the probability sum rules in the nonunitarity form


\begin{equation}
\sum_\beta P(\nu_\alpha \rightarrow \nu_\beta) = \sum_i | V_{i \alpha}|^2 
e^{-2 \gamma_i t} \;,
\end{equation}


\ni in spite of the unitarity of $ V $. Of course, the rhs of Eq. (92) is 
equal to 1, if all (here involved) $\gamma_i $ are zero. In this case, the 
damping in Eqs. (91) disappears and they become the {\it conventional} neutrino
oscillation formulae. The same is true for the next Eqs. (93).

 If the quartic products in Eqs. (91) are real (as it turns out to be in our 
case), we can rewrite these equations in the form


\begin{eqnarray}
P(\nu_\alpha \rightarrow \nu_\beta) & = & \sum_{j\,i}V_{j \beta}V_{j \alpha}^* 
V_{i \beta}^* V_{i \alpha} e^{-(\gamma_j + \gamma_i)t} \nonumber \\ & - & 
\sum_{j>i} V_{j \beta}V_{j \alpha}^* V_{i \beta}^* V_{i \alpha} \sin^2\left(
\frac{E_j - E_i}{2}\,t\right) e^{-(\gamma_j + \gamma_i)t} \;,
\end{eqnarray}


\ni where the first term is equal to


\begin{equation}
\delta_{\beta \alpha} - \sum_{j\,i} V_{j \beta}V_{j \alpha}^* V_{i \beta}^* 
V_{i \alpha} \left[1 - e^{-(\gamma_j + \gamma_i)t} \right]\;.
\end{equation}


\ni Writing $ (E_j - E_i)t = \Delta m^2_{j\,i} L/2E $ and $(\gamma_j + 
\gamma_i)t = (|m_{\nu_j}|\gamma^{(0)}_j + |m_{\nu_i}|\gamma^{(0)}_i) L/E $ with
$\Delta m^2_{j\,i} \equiv  m^2_{\nu_j} - m^2_{\nu_i} $, $ E = |\vec{p}|$ and $
L = t $, and then expressing the neutrino masses $ m_{\nu_i}$ and rest widths
$\gamma^{(0)}_i $ in eV, the experimental baseline $ L $ in km and the neutrino
beam energy in GeV, we can insert


\begin{eqnarray}
\frac{E_j - E_i}{2}\,t & \rightarrow & 1.27 \frac{\Delta m^2_{j\,i} L}{E}
\equiv x_j - x_i \;,\nonumber \\ (\gamma_j + \gamma_i)t & \rightarrow & 5.07 
\frac{(|m_{\nu_j}|\gamma^{(0)}_j + |m_{\nu_i}|\gamma^{(0)}_i) L}{E} \equiv y_j
+ y_i
\end{eqnarray}

\vspace{0.1cm}

\ni in Eq. (91) and (93) (here, $ c = 1 = \hbar $){\footnote{The insertion $ L
= vt $ with $ v = |\vec{p}|/E \simeq c\;\;(c = 1)$ is called by Lipkin [10]
the "right handwaving" which converts the "gedanken oscillation experiment" 
{\it in time} into the real oscillation experiment {\it in space}. In the first
experiment, a flavor neutrino is created by a weak--interaction source (of 
size $\ll L $) in a momentum eigenstate $|\nu_\alpha, \vec{p}\rangle $ being a 
superposition of a few energy eigenstates $|\nu_i, E_i \rangle $ (with $ E_i = 
\sqrt{\vec{p}^{\,2}+ m^2_{\nu_i}}$) describing mass neutrinos evolving in time.
Inversely, in the second experiment, the flavor neutrino is emitted in an 
energy eigenstate $|\nu_\alpha, E \rangle $ given as a superposition of a few 
momentum eigenstates $|\nu_i, \vec{p}_i\rangle $ (with $|\vec{p}_i| = \sqrt{
E^2 - m^2_{\nu_i}}$) describing mass neutrinos propagating in space (the 
requirement of coherence within this superposition leads to the condition $|\,|
\vec{p}_i| - |\vec{p}_j|\,| \ll 1/{\rm source\;size} $). In the first case $ 
E_i - E_j \simeq \Delta m^2_{ij}/2|\vec{p}| $, while in the second $ |\vec{p}_i
| - |\vec{p}_j| \simeq \Delta m^2_{ij}/2 E $. Here, $ E \simeq c |\vec{p}|\;\;
(c = 1)$. A "wrong handwaving" would be the insertion $ L = v_i t_i $ with $ 
v_i = \vec{p}/ E_i $.}}. 

 From Eqs. (93) with $ V_{i \alpha} = U_{\alpha i}^{(\nu)\,*}$, we derive in 
the case of our form (84) of $ U_{\alpha i}^{(\nu)}$ the following damped 
oscillation formulae for active neutrinos $\nu_e\,,\,\nu_\mu\,,\,\nu_\tau $
(in the vacuum): 

\vspace{0.1cm}

\begin{eqnarray}
P(\nu_e \rightarrow \nu_\mu) & = & 0 = P(\nu_\mu \rightarrow \nu_e) \;, 
\nonumber \\  P(\nu_e \rightarrow \nu_\tau) & = & 0 = P(\nu_\tau \rightarrow 
\nu_e) \;, \nonumber \\ P(\nu_\mu \rightarrow \nu_\tau) & = & 0 = P(\nu_\tau 
\rightarrow \nu_\mu)\;, \nonumber \\ 
P(\nu_e \rightarrow \nu_e) & = & \left(\frac{e^{-y_1} + X^2e^{-y_4}}{1 + X^2}
\right)^2 - \left(\frac{2 X}{1 + X^2}\right)^2 \sin^2(x_4 - x_1) e^{-(y_4 + 
y_1)} \;,\nonumber \\ 
P(\nu_\mu \rightarrow \nu_\mu) & = & \left(\frac{e^{-y_2} + Y^2 e^{-y_5}}{1 + 
Y^2}\right)^2  - \left(\frac{2 Y}{1 + Y^2}\right)^2 \sin^2(x_5 - x_2) e^{-
(y_5 + y_2)}\;, \nonumber \\ P(\nu_\tau \rightarrow \nu_\tau) & = & e^{-2y_3}
\end{eqnarray}

\vspace{0.1cm}

\ni and those where, beside $\nu_e\,,\,\nu_\mu\,,\,\nu_\tau $, the sterile 
neutrinos $\enu\,,\,\mnu $ participate explicitly:

\vspace{0.1cm}

\begin{eqnarray}
P(\nu_e \rightarrow \enu) & = & \left(\frac{X (e^{-y_1} - e^{-y_4})}{1 + X^2}
\right)^2 + \left(\frac{2 X}{1 + X^2}\right)^2 \sin^2(x_4 - x_1) e^{-(y_4 + 
y_1)} \;, \nonumber \\  P(\nu_e \rightarrow \mnu) & = & 0 \;, \nonumber \\ 
P(\nu_\mu \rightarrow \enu) & = & 0 \;, \nonumber \\ 
P(\nu_\mu \rightarrow \mnu) & = & \left(\frac{Y (e^{-y_2} -e^{-y_5})}{1 + Y^2}
\right)^2 + \left(\frac{2 Y}{1 + Y^2}\right)^2 \sin^2(x_5 - x_2) e^{-(y_5 + 
y_2)} \;,\nonumber \\ P(\nu_\tau \rightarrow \enu) & = & 0 \;, \nonumber \\ 
P(\nu_\tau \rightarrow \mnu) & = & 0 \;.
\end{eqnarray}

\vspace{0.1cm}

\ni The probabilities (96) and (97) satisfy the sum rules (92) which now read :

\vspace{0.1cm}

\begin{eqnarray}
P(\nu_e \rightarrow \nu_e) + P(\nu_e \rightarrow \enu) & = & \frac{e^{-2y_1} + 
X^2 e^{-2y_4}}{1 + X^2} \;, \nonumber \\ 
P(\nu_\mu \rightarrow \nu_\mu) + P(\nu_\mu \rightarrow \mnu) & = & \frac{
e^{-2y_2} + Y^2 e^{-2y_5}}{1 + Y^2} \;.
\end{eqnarray}

\vspace{0.1cm}

 Note that damping in our neutrino oscillation formulae decreases with growing 
neutrino energy $ E $, because $ y_i = 5.07 |m_{\nu_i}| \gamma^{(0)}_i L/E $ 
decreases. Thus, the larger $\nu_\alpha$--neutrino energy is explored in
$\nu_\alpha$--neutrino experiments, the smaller damping influence is exerted on
$P(\nu_\alpha \rightarrow \nu_\alpha)$, provided not all (involved) $\gamma_i $
are zero. Of course, the effect of damping, if any, is expected to be very 
small. 

\vfill\eject

\vspace{0.3cm}

\ni {\bf 7. A mechanism of negligible damping}

\vspace{0.3cm}

 Now, we turn to the discussion of a possible mechanism of neutrino instability
{\it i.e.,} instability of mass neutrino states. To this end observe that the
neutrino weak current

\begin{equation}
J^{(\nu)\,\mu} = \overline{\nu_{eL}}\gamma^\mu \nu_{eL} + \overline{\nu_{\mu L
}} \gamma^\mu \nu_{\mu L} + \overline{\nu_{\tau L}} \gamma^\mu \nu_{\tau L}\;,
\end{equation}

\ni though it is diagonal in the active neutrinos $\nu_e\,,\,\nu_\mu\,,\,
\nu_\tau $, is no longer diagonal in the mass neutrinos $\nu_1\,,\,\nu_2\,,\,
\nu_3\,,\,\nu_4\,,\,\nu_5 $, if the sterile neutrinos $\enu $, $\mnu $ really
exist. In fact, inserting in Eq. (99) the unitary transformation (86), we 
obtain generally, beside $\overline{\nu_{iL}}\gamma^\mu \nu_{iL}$, some non\-%
diagonal products $\overline{\nu_{iL}}\gamma^\mu \nu_{jL}\;\;(i \neq j)$, since
only three of five products $\overline{\nu_{\alpha L}}\gamma^\mu \nu_{\alpha L}
$ are originally present in Eq. (99).

 For instance, in the case of our form (84) of $ U_{\alpha i}^{(\nu)}$, the 
unitary transformation (86) with $ V_{i \alpha}^* = U_{\alpha i}^{(\nu)}$ gives

\vspace{-0.2cm}

\begin{eqnarray}
\nu_e & = & \frac{1}{\sqrt{1 + X^2}}\,\left(\nu_1 - X \nu_4 e^{i \varphi^{(\nu
)}}\right)\;, \nonumber \\ \nu_\mu & = & \frac{1}{\sqrt{1 + Y^2}}\,\left(\nu_2 
- Y \nu_5 e^{i \varphi^{(\nu)}}\right)\;, \nonumber \\ \nu_\tau & = & \nu_3 \;,
\nonumber \\ \enu & = & \frac{1}{\sqrt{1 + X^2}}\,\left(X \nu_1 e^{-i 
\varphi^{(\nu)}} + \nu_4 \right)\;, \nonumber \\ \mnu & = & \frac{1}{\sqrt{1 + 
Y^2}}\,\left(Y \nu_2 e^{-i \varphi^{(\nu)}} + \nu_5 \right)\;.
\end{eqnarray}

\ni Thus, in our case, the neutrino weak current (93) transits into the form 

\begin{eqnarray}
\!\!J^{(\nu)\,\mu} & = & \frac{1}{1 + X^2}\left[\overline{\nu_{1L}}\gamma^\mu 
\nu_{1L} + X^2 \overline{\nu_{4L}}\gamma^\mu \nu_{4L} - X\left(\overline{\nu_{
1L}} \gamma^\mu \nu_{4L}e^{\varphi^{(\nu)}} + \overline{\nu_{4L}}\gamma^\mu 
\nu_{1L}e^{-i \varphi^{(\nu)}}\right) \right] \nonumber \\ & & +\; 
\overline{\nu_{3L}}\gamma^\mu \nu_{3L} \nonumber \\ & & + \;\frac{1}{1 + Y^2}
\left[\overline{\nu_{2L}}\gamma^\mu \nu_{2L} + Y^2 \overline{\nu_{5L}}
\gamma^\mu \nu_{5L} - Y\left(\overline{\nu_{2L}}\gamma^\mu \nu_{5L}e^{\varphi^{
(\nu)}} + \overline{\nu_{5L}}\gamma^\mu \nu_{2L}e^{-i \varphi^{(\nu)}}
\right)\right] \;.\nonumber \\ & &
\end{eqnarray}

\ni Since in the Standard Model lagrangian this neutrino weak current is 
coupled to the $ Z $ boson [with the coupling constant $ -g/(2\cos \theta_W) = 
- e/(2\sin \theta_W \cos \theta_W) $], some neutrino decays of the type $\nu_i 
\rightarrow \nu_j\,\nu_k\,\bar{\nu_l} $ with $(i,j) = (1,4)$ or $(4,1)$ and 
$(2,5)$ or $(5,2)$, and with similar $(k,l)$, are $ Z $--mediated, so that 
they can be real processes if only $|m_{\nu_i}| > |m_{\nu_j}| + |m_{\nu_k}| + 
|m_{\nu_l}| $ (here, $\bar{\nu}_l $ denotes an antiparticle of $\nu_l $).

 In the case of our neutrino mass spectrum (83), we get the inequalities $
m_{\nu_1} > |m_{\nu_4}|$, $ m_{\nu_2} > |m_{\nu_5}|$ and $ m_{\nu_2} > m_{
\nu_1}$, where in the last relation we make use of $ M^{(\nu)}_{22}>M^{(\nu
)}_{11}$. Further, $|m_{\nu_5}| > m_{\nu_1}$, $|m_{\nu_5}| > |m_{\nu_4}|$, 
$m_{\nu_3} > m_{\nu_2}$ and $ m_{\nu_3} > |m_{\nu_5}|$, if $ Y - X > M^{(\nu
)}_{11}/|M^{(\nu)}_{25}|$, $ Y > X $, $ Y < (M^{(\nu)}_{33} - M^{(\nu)}_{22})/
|M^{(\nu)}_{25}|$ and $ Y < M^{(\nu)}_{33}/|M^{(\nu)}_{25}|$, respectively. 
Thus, for $ Y - X > M^{(\nu)}_{11}/|M^{(\nu)}_{25}|$ and $ Y < (M^{(\nu)}_{33} 
- M^{(\nu)}_{22})/|M^{(\nu)}_{25}|$ all these inequalities hold. In 
this case, therefore,

\begin{equation}
|m_{\nu_4}| < m_{\nu_1} < |m_{\nu_5}| < m_{\nu_2} < m_{\nu_3} \;,
\end{equation}

\ni showing that then $|m_{\nu_4}| $ is the lowest neutrino mass.

 We can see that for any virtual decay $\nu_1 \rightarrow \nu_4\,\nu_k\,
\bar{\nu}_l $ we get 

\begin{eqnarray}
m_{\nu_1} - |m_{\nu_4}| - |m_{\nu_k}| - |m_{\nu_l}| & \leq & m_{\nu_1} - 
|m_{\nu_4}| - 2|m_{\nu_4}| = 2M^{(\nu)}_{11} - \sqrt{M^{(\nu)\,2}_{11} + 4 
|M^{(\nu)}_{14}|^2} \nonumber \\ & = & M^{(\nu)}_{11} - 2|M^{(\nu)}_{14}| X 
>\;\,{\rm or}\;\,\leq 0 \,,
\end{eqnarray}

\ni depending on $ X < $ or $\geq M^{(\nu)}_{11}/2|M^{(\nu)}_{14}| $. This 
implies that, {\it a priori}, the decay width of $\nu_1 $ neutrino may be $
\gamma_1 \neq 0 $ or $\gamma_1 = 0 $, respectively. Since $|m_{\nu_4}| < 
m_{\nu_1}$, no virtual decay $\nu_4 \rightarrow \nu_1\,\nu_k\,\bar{\nu_l}$ can 
be a real process, what leads to $\gamma_4 = 0 $ for $\nu_4 $ neutrino.

Similarly, for any virtual decay $\nu_2 \rightarrow \nu_5\,\nu_k\,\bar{\nu_l}$,
we obtain

\begin{eqnarray}
m_{\nu_2}\! -\! |m_{\nu_5}|\! -\! |m_{\nu_k}|\! -\! |m_{\nu_l}| & \leq & 
m_{\nu_2}\! -\! |m_{\nu_5}|\! -\! 2|m_{\nu_4}| = M^{(\nu)}_{11} + M^{(\nu)}_{
22}\! -\! \sqrt{M^{(\nu)\,2}_{11} + 4 |M^{(\nu)}_{14}|^2} \nonumber \\ & = & 
M^{(\nu)}_{22}\! -\! 2|M^{(\nu)}_{14}| X > 0 
\end{eqnarray}

\ni if $X < M^{(\nu)}_{22}/2|M^{(\nu)}_{14}|$, where $ M^{(\nu)}_{22} = 
(4/9)(80/\varepsilon^{(\nu)} - 1)M^{(\nu)}_{11}$ with $\varepsilon^{(\nu)} < 
1 $ ({\it cf.} Eq. (1) with $ f = \nu $). If true, this gives a nonzero
decay width $\gamma_2 \neq 0 $ for $\nu_2 $ neutrino. On the other hand, for 
$\nu_5 $ neutrino $\gamma_5 = 0 $, since $|m_{\nu_5}| < m_{\nu_2}$.

 Anticipating that $\gamma_1 = 0$ (or is extremely small) and putting $\gamma_3
= \gamma_4 = \gamma_5 = 0 $, we obtain from Eqs. (96) and (97) the following 
neutrino oscillation formulae (possibly damped if $\gamma_2 \neq 0 $):

\begin{eqnarray}
\!P(\nu_e \rightarrow \nu_e) &\!\! = & \!\!1 - \left(\frac{2X}{1 + X^2}\right)^{
\!2} \sin^2(x_4 - x_1) = 1 -  P(\nu_e \rightarrow \enu) \;, \nonumber \\ 
\!P(\nu_\mu \rightarrow \nu_\mu) &\!\! = & \!\!\left(\frac{e^{-y_2} + Y^2}{1 + 
Y^2}\right)^{\!2} - \left(\frac{2Y}{1 + Y^2}\right)^{\!2} \sin^2(x_5 - x_2) = 
\frac{e^{-2y_2} + Y^2}{1 + Y^2} - P(\nu_\mu \rightarrow \mnu) \;, \nonumber \\ 
\!P(\nu_\tau \rightarrow \nu_\tau) &\!\! = & \!\!1 \;. 
\end{eqnarray}

\ni Here,

\begin{equation}
x_1 - x_4 = 2.53 \frac{|M^{(\nu)}_{14}|\, M^{(\nu)}_{11} L}{E}\;\;,\;\; x_2 - 
x_5 = 2.53 \frac{|M^{(\nu)}_{25}|\, M^{(\nu)}_{22} L}{E}\;.
\end{equation}

 From the neutrino mass spectrum (83) and the definitions (85) of $ X $ and $ Y
$, we can derive the useful equations expressing $ M^{(\nu)}_{11}$ and $ |M^{(
\nu)}_{14}|$ through $ X $ and $\Delta m^2_{14}$, as well as $ M^{(\nu)}_{22}$ 
and $|M^{(\nu)}_{25}|$ through $ Y $ and $\Delta m^2_{25}$:

\begin{equation}
M^{(\nu)}_{11} = \left(\frac{1-X^2}{1+X^2}\Delta m^2_{14}\right)^{1/2}\;,\;\; 
|M^{(\nu)}_{14}| = \left(\frac{X^2}{1 - X^4}\Delta m^2_{14}\right)^{1/2}
\end{equation}

\ni as well as

\begin{equation}
M^{(\nu)}_{22} = \left(\frac{1-Y^2}{1+Y^2}\Delta m^2_{25}\right)^{1/2}\;,\;\; 
|M^{(\nu)}_{25}| = \left(\frac{Y^2}{1 - Y^4}\Delta m^2_{25}\right)^{1/2}\;.
\end{equation}

\ni Further, writing 

\begin{equation}
1 \geq \left(\frac{2 X}{1+X^2}\right)^{\!2} \equiv \sin^2 2\theta^{(e)}\;\;,\;\;
1 \geq \left(\frac{2 Y}{1+Y^2}\right)^{\!2} \equiv \sin^2 2\theta^{(\mu)}\;,
\end{equation}

\ni we obtain

\begin{equation}
1 \geq X \equiv \tan \theta^{(e)}\;\;,\;\;1 \geq Y \equiv \tan \theta^{(\mu)}
\;,
\end{equation}

\ni where $ 0 \leq 2\theta^{(e)} \leq \pi/2 $ and $ 0 \leq 2\theta^{(\mu)} 
\leq \pi/2 $. We can see from Eqs. (108) that for a fixed finite $|M^{(\nu)}_{
25}|$ we get $\Delta m^2_{25} \rightarrow 0 $ as $ Y \rightarrow 1$, excluding
in this limit the corresponding neutrino oscillations. On the other hand, if
we insist in an argument to keep $\Delta m^2_{25}$ fixed and nonzero as $ Y 
\rightarrow 1 $, we formally have $|M^{(\nu)}_{25}| \rightarrow \infty$, imply%
ing $ m_{\nu_2} \rightarrow \infty $ and $|m_{\nu_5}| \rightarrow \infty $. (In
both cases $ M^{(\nu)}_{22} \rightarrow 0 $ as $ Y \rightarrow 1 $.) Analogical
conclusions follow from Eqs. (107) for $|M^{(\nu)}_{14}|$ and $\Delta m^2_{14}$
(and $ M^{(\nu)}_{11} $) when $ X \rightarrow 1 $.

 The first Eq. (105) enables us to ascribe the observed deficit of solar $\nu_e
$'s to $\nu_e \rightarrow \enu $ oscillations. In fact, we can determine our 
parameters $ M^{(\nu)}_{11}$ and $ |M^{(\nu)}_{14}|$ putting 

\begin{equation}
\left(\frac{2 X}{1+X^2}\right)^{\!2} = \sin^2 2\theta_{\rm solar} \sim 0.75\;\;,\;\;
\Delta m^2_{14} = \Delta m^2_{\rm solar} \sim 6.5\times 10^{-11}\;{\rm eV}^2
\;,
\end{equation}

\ni if the global vacuum fit to solar data [5] is chosen. Then, due to Eqs. 
(110) and (107)

\begin{equation}
X = \tan \theta_{\rm solar} \sim 1/\sqrt{3} = 0.577\;\;,\;\;M^{(\nu)}_{
11} \sim 5.70 \times 10^{-6}\;{\rm eV}\;\;,\;\;|M^{(\nu)}_{14}| \sim 4.94
\times 10^{-6}\;{\rm eV}\;.
\end{equation}

\ni Here, we can see that $ M^{(\nu)}_{11}/2|M^{(\nu)}_{14}| = (1 - X^2)/2X 
\sim 1/\sqrt{3} \sim X $. Thus, the condition leading to $\gamma_1 = 0 $ is 
satisfied on the edge [{\it cf.} Eq. (103)]. At the same time, this shows that
the condition $ M^{(\nu)}_{22}/2|M^{(\nu)}_{14}| > X $, providing $\gamma_2 
\neq 0 $ in the second Eq. (105), is fulfilled comfortably [{\it cf.} Eq. 
(104)].

 Damping in the second Eq. (105)  complicates our discussion, though it is 
natural to expect that this formula allows us to ascribe the observed deficit 
of atmospheric $\nu_\mu$'s to $\nu_\mu \rightarrow \mnu $ oscillations. In 
fact, anticipating that damping in this case is tiny [{\it cf.} Eq. (119)], 
we may write $\exp(-y_2) \simeq 1 - y_2 $ and, therefore,

\begin{equation}
P(\nu_\mu \rightarrow \nu_\mu) \simeq 1 - \left(\frac{2Y}{1 + Y^2}\right)^{\!
2} \sin^2(x_5 - x_2) - y_2 \left(\frac{2Y}{1 + Y^2}\right)^{\!2}\left[
\frac{1}{2} - \sin^2(x_5 - x_2)\right]\;,
\end{equation}

\ni where the coefficient at $ y_2 $ in the correction $ O(y_2)$ is almost 
compensated to zero. Thus, we can put approximately

\begin{equation}
\left(\frac{2 Y}{1+Y^2}\right)^{\!2} \simeq \sin^2 2\theta_{\rm atm} \sim 
0.82\;\;{\rm to}\;\;1\;\;,\;\;\Delta m^2_{25} \simeq \Delta m^2_{\rm atm} \sim 
(0.5\;\;{\rm to}\;\;6)\times 10^{-3}\;{\rm eV}^2\;,
\end{equation}

\ni where the recent data from \UK atmospheric neutrino experiment [4] is 
applied. Here, we will put, for instance, $\sin^2 2\theta_{\rm atm} \sim 0.999 
$ and $\Delta m^2_{\rm atm} \sim 5\times 10^{-3}$ eV$^2$ as in Section~3. Then,

\begin{equation}
Y \simeq \tan \theta_{\rm atm} \sim 0.969\;\;,\;\;M^{(\nu)}_{22} \sim 0.126
\times 10^{-1}\,{\rm eV}\;\;,\;\;|M^{(\nu)}_{25}| \sim 1.99\times 10^{-1}{\rm 
eV}
\end{equation}

\ni due to Eqs (110) and (108).

 Making use of the estimations (112) and (115), we can evaluate $\varepsilon^{(
\nu)}$ and $\mu^{(\nu)}$ from Eq. (1) (with $ f = \nu$),

\begin{eqnarray}
\varepsilon^{(\nu)} & = & \frac{80}{1 + 9 M^{(\nu)}_{22}/ 4 M^{(\nu)}_{11}}
\sim 1.61\times 10^{-2} \;,\nonumber \\
\mu^{(\nu)} & = & \frac{29 M^{(\nu)}_{11}}{\varepsilon^{(\nu)}} \sim 1.03\times
10^{-2}\;{\rm eV}\;, 
\end{eqnarray}

\ni and then, the neutrino masses $ m_{\nu_1}$, $ m_{\nu_4}$, $ m_{\nu_2}$, 
$ m_{\nu_5}$ and $ m_{\nu_3}$ from Eqs. (83),

\begin{eqnarray}
m_{\nu_1} & \sim & 8.55\times 10^{-6}\;{\rm eV}\;\;,\;\;m_{\nu_4} \sim 
-2.85\times 10^{-6}\;{\rm eV} \;,\nonumber \\
m_{\nu_2} & \sim & 2.05\times 10^{-1}\;{\rm eV}\;\;,\;\;m_{\nu_5} \sim 
-1.93\times 10^{-1}\;{\rm eV}
\end{eqnarray}

\ni and

\begin{equation}
m_{\nu_3} = \frac{\mu^{(\nu)}}{29} \frac{24}{25}\left(624 + \varepsilon^{(\nu)
}\right) \sim 2.12\times 10^{-1}\;{\rm eV}\;.
\end{equation}

\ni These masses satisfy consistently the inequalities (102) and reproduce the
experimental values (101) and (114): $\Delta m^2_{14} \sim 6.5\times 10^{-11}
\;{\rm eV}^2 $ and $\Delta m^2_{25} \sim 5\times 10^{-3}\;{\rm eV}^2 $.

 Now, we can evaluate the total decay width at rest, $\gamma^{(0)}_i $, for a 
mass neutrino $\nu_i$ decaying through the $ Z $--mediated processes $\nu_i
\rightarrow \nu_j\,\nu_k\,\bar{\nu}_l $, where $ m_i = E_j + E_k + E_l > m_j
+ m_k + m_l $ with $ m_n = |m_{\nu_n}| $. In the case of $ m_2 $, $ m_5 $ and
$ m_2 - m_5 $ dominating over $ m_k $ and $ m_l $ ($ k,\,l = 1, 4$), we obtain 
the approximate formula

\begin{equation}
\gamma^{(0)}_2 = \frac{1}{4}\, \frac{ G^2_F}{192 \pi^3} \left(\frac{Y}{1+Y^2}
\right)^{\!2}\left(m_2 - m_5\right)^4\left(m_2 + 2 m_5\right)\;,
\end{equation}

\ni where the total decay width $\gamma^{(0)}_2 $ is the sum of four partial 
decay widths for $\nu_2 \rightarrow \nu_5\,\nu_k\,\bar{\nu}_l $ with $(k\,,\,l)
= (1\,,\,4)\,,\,(4\,,\,1)\,,\,(1\,,\,1)\,,\,(4\,,\,4)$ which are proportional
to

\vspace{-0.6cm}

$$
\!\left(\frac{Y}{1+Y^2}\right)^{\!2}\!\!\!\left(\frac{X}{1+X^2}\right)^{\!2}
\!,\,\left(\frac{Y}{1+Y^2}\right)^{\!2}\!\!\!\left(\frac{X}{1+X^2}\right)^{\!2}
\!,\,\left(\frac{Y}{1+Y^2}\right)^{\!2}\!\!\!\left(\frac{1}{1+X^2}\right)^{\!2}
\!,\,\left(\frac{Y}{1+Y^2}\right)^{\!2}\!\!\!\left(\frac{X^2}{1+X^2}\right)^{
\!2}\!,
$$

\ni respectively, the sum of these weights being equal to $Y^2/(1+Y^2)^2$. In 
this calculation, we used the Standard Model coupling of the neutrino weak 
current (101) to the $ Z $ boson [with the coupling constant $ -g/(2\cos 
\theta_W)$, where $ G_F/\sqrt{2} = g^2/(8 M_W) = g^2/(8 M_Z \cos \theta_W )$], 
and considered the situation when $(p_2 - p_5) \ll M^2_Z $ at the rest frame 
of decaying $\nu_2 $: $p_2 = (m_2\,,\,\vec{0})$. In Eq. (119), the factor 1/4 
at the front is a consequence of using the neutral weak current (rather than 
charged weak current), while $Y^2/(1+Y^2)^2$ stems from mixing of active and 
sterile neutrinos.

 If $ Y $, $ m_2 $ and $ m_5 $ are estimated as in Eqs. (115) and (117), then 
the formula (119) gives (with the Fermi constant $ G_F = 1.17\times 10^{-5}\;
{\rm GeV}^{-2}$) the extremely small value 

\begin{equation}
\gamma^{(0)}_2 \sim 10^{-59}\;{\rm eV}
\end{equation}

\ni corresponding to the enormous lifetime $\tau_2 = 1/\gamma^{(0)}_2 \sim 
10^{43}$ sec (as ${\rm eV}^{-1} = 6.58\times 10^{-16}$  sec). This implies for 
the \UK atmospheric experiment that $ y_2 = 5.07 m_2\gamma^{(0)}_2 L/E $ $\sim 
10^{-55}$ with $ m_2\gamma^{(0)}_2 \sim 10^{-60}$, $L \sim 1.3\times 10^4 $ and
$ E \sim 1 $ expressed in eV$^2$, km and GeV, respectively. Thus, practically, 
$ y_2 = 0 $ and so $\exp(-y_2) = 1 $. If $m_2 = m_{\nu_2}$ and $m_5 = |m_{
\nu_5}|$ grow by one order of magnitude (what is the case when $\sin 2
\theta_{\rm atm}$ rises to 0.9999 and so, $ Y $~to~ 0.990), then $\gamma^{(0
)}_2 $ becomes not larger than $ \sim 10^{-54}$ eV and $\tau_2 $ not smaller 
than $\sim 10^{38}$ sec.

 Concluding the last Section, we can say that damping in neutrino oscillation 
formulae can be completely neglected, unless there are {\it other} sources of 
neutrino instability [11], more {\it effective} than the $ Z $--mediated decays
$\nu_i \rightarrow \nu_j\,\nu_k\,\bar{\nu}_l $ considered in this paper. The 
last decays appear in the Standard Model framework if, additionally, there are
sterile neutrinos mixing with the active ones and so, breaking the elektroweak 
symmetry $ SU(2)\times U(1)$. Our discussion shows that the neutrino decay 
widths $\gamma_i $ are zero for $ i = 1,3,4,5 $ and are completely negligible 
for $ i = 2 $. However, our damped oscillation formulae (93) [and their more 
specific versions given in Eqs. (96) and (97)] can work for any sort of 
potential neutrino instability.

\vfill\eject

\vspace{0.3cm}

{\centerline{\bf Appendix: Majorana sterile neutrinos}}

\vspace{0.3cm}


 The flavor neutrinos, three active $\nu_e\,,\,\nu_\mu\,,\,\nu_\tau $ and two 
sterile $\enu $, $\mnu $, considered in Sections 5, 6 and 7, lead to five mass 
neutrinos $\nu_1\,,\,\nu_2\,,\,\nu_3\,,\,\nu_4\,,\,\nu_5 $ having pure Dirac 
masses (also in previous Sections neutrinos had always pure Dirac masses). Now,
assume that there are {\it solely} three active flavor neutrinos, {\it but} 
they possess the "Majorana" $ 2\times 2 $ mass matrices

\vspace{-0.1cm}

$$
\widehat{M}^{(\nu)}_{\alpha} = \left( \begin{array}{cc} m^{(L)}_\alpha & 
m^{(D)}_\alpha \\ m^{(D)}_\alpha & m^{(R)}_\alpha \end{array} \right)\;\;
(\alpha = e\,,\,\mu\,,\,\tau)\;\;,
\eqno({\rm A}.1)
$$

\ni each consisting of {\it one} Dirac and {\it two} Majorana masses, $ m^{(D)
}_\alpha $ and $ m^{(L,R)}_\alpha $, respectively [12]. The mass matrices (A.1)
imply the following mass term in the lagrangian:

\vspace{-0.1cm}

$$
-{\cal L}_{\rm mass} = \frac{1}{2} \sum_\alpha \left( \overline{\nu^{(a)
}_\alpha}\;\;\;\; \overline{\nu^{(s)}_\alpha} \right)
\;\widehat{M}^{(\nu)}_\alpha \;\left(\begin{array}{c} \nu^{(a)}_\alpha \\ 
\nu^{(s)}_\alpha \end{array} \right)\;\;,
\eqno({\rm A}.2)
$$

\ni where

\vspace{-0.2cm}

$$
\nu^{(a)}_\alpha \equiv \nu_{\alpha\,L} + \left(\nu_{\alpha\,L}\right)^c\,,\,
\nu^{(s)}_\alpha \equiv \nu_{\alpha\,R} + \left(\nu_{\alpha\,R}\right)^c
\;\; (\alpha = e\,,\,\mu\,,\,\tau)
\eqno({\rm A}.3)
$$

\ni are the Majorana flavor neutrinos, three active $\nu^{(a)}_\alpha $ and 
three sterile $\nu^{(s)}_\alpha $, built up of chiral fields $\nu_{\alpha\,L}$,
$\left(\nu_{\alpha\,L}\right)^c = \left(\nu_{\alpha}\right)^c_R $ and $\nu_{
\alpha\,R}$, $\left(\nu_{\alpha\,R}\right)^c = \left(\nu_{\alpha}\right)^c_L $ 
involved already in the Dirac flavor neutrinos $\nu_\alpha = \nu_{\alpha\,L} +
\nu_{\alpha\,R}$ and antineutrinos $\nu_\alpha^c = (\nu_{\alpha\,L})^c + (\nu_{
\alpha\,R})^c $. These {\it conventional} Majorana sterile neutrinos $\nu^{(s)
}_\alpha $ contain, therefore, {\it no extra} neutrino degrees of freedom, in 
contrast to our previous Dirac sterile neutrinos $\nu^{(e,\mu)}_{s} = \nu^{(e,
\mu)}_{s\,L} + \nu^{(e,\mu)}_{s\,R}$ involving {\it extra} chiral fields $\nu^{
(e,\mu)}_{s\,L}$ and $\nu^{(e,\mu)}_{s\,R}$. Of course, in contrast to the 
Dirac, the Majorana neutrinos mix (maximally) the lepton number $ L $.

 In the case of "Majorana" mass matrices (A.1), the overall neutrino mass 
matrix takes the $ 6\times 6 $ form

\vspace{-0.1cm}

$$
\widehat{M}^{(\nu)} = \left(\delta_{\alpha \beta}\widehat{M}^{(\nu)}_{\alpha}
\right) = \left(\delta_{\alpha \beta}\left( \begin{array}{cc} m^{(L)}_\alpha & 
m^{(D)}_\alpha \\ m^{(D)}_\alpha & m^{(R)}_\alpha \end{array} \right)\right)\;
\;. \eqno({\rm A}.4)
$$

\ni In this "pure--Majorana" mass matrix there is no mixing between flavor 
neutrinos from three lepton families $\alpha = e\,,\,\mu\,,\,\tau $.

 Diagonalizing the "pure--Majorana" mass matrix (A.4), we obtain the neutrino 
masses

$$
m^{I,\,II}_{\alpha} = \frac{m^{(L)}_\alpha + m^{(R)}_\alpha}{2} \mp \sqrt{
\left( \frac{m^{(L)}_\alpha + m^{(R)}_\alpha}{2}\right)^2 + m^{(D)\,2}_\alpha} 
\simeq \frac{m^{(L)}_\alpha + m^{(R)}_\alpha}{2} \mp  m^{(D)}_\alpha 
\eqno({\rm A}.5)
$$

\ni corresponding to six Majorana mass neutrinos

\vspace{-0.1cm}

\begin{eqnarray*}
\nu_\alpha^{I} & = & \cos \theta_\alpha \nu_\alpha^{(a)} - \sin \theta_\alpha
\nu_\alpha^{(s)}\;\;,\\
\nu_\alpha^{II} & = & \sin \theta_\alpha \nu_\alpha^{(a)} + \cos \theta_\alpha
\nu_\alpha^{(s)}\;\;,
\end{eqnarray*}

\vspace{-1.55cm}

\begin{flushright}
(A.6)
\end{flushright}

\vspace{0.07cm}

\ni where

\vspace{-0.2cm}

\begin{eqnarray*}
\cos \theta_\alpha & = & \frac{m^{(D)}_{\alpha}}{\sqrt{m^{(D)\,2}_\alpha + 
\left(m^{II}_\alpha - m^{(R)}_\alpha\right)^2}} \simeq \frac{1}{\sqrt{2}}\left(
1 - \frac{m^{(L)}_\alpha - m^{(R)}_\alpha}{4 m^{(D)}_\alpha}\right) \simeq 
\frac{1}{\sqrt{2}}\;, \\
\sin \theta_\alpha & = & \frac{m^{II}_{\alpha} -m^{(R)}_{\alpha}}{\sqrt{m^{(D)
\,2}_\alpha + \left(m^{II}_\alpha - m^{(R)}_\alpha \right)^2}} \simeq 
\frac{1}{\sqrt{2}}\left( 1 + \frac{m^{(L)}_\alpha - m^{(R)}_\alpha}{4 m^{(D)
}_\alpha}\right) \simeq \frac{1}{\sqrt{2}}
\end{eqnarray*}

\vspace{-1.92cm}

\begin{flushright}
(A.7)
\end{flushright}

\vspace{0.4cm}

\ni with $ \theta_\alpha \simeq \pi/4 + \left(m^{(L)}_\alpha - m^{(R)}_\alpha\right)
/4 m^{(D)}_\alpha \simeq \pi/4 $ ($ m^{I}_\alpha$ may be negative). Here, 
the approximate equalities are valid in the case of $ m^{(L)}_\alpha \simeq 
m^{(R)}_\alpha $. If in addition $ m^{(L)}_\alpha \simeq m^{(R)}_\alpha \simeq 
m^{(D)}_\alpha $, then Eqs. (A.5) give $ m^{I}_\alpha \simeq 0 $ and $ m^{II
}_\alpha \simeq 2m^{(D)}_\alpha $. In contrast, if $ m^{(L)}_\alpha \simeq 
m^{(R)}_\alpha \ll m^{(D)}_\alpha $, they imply $ m^{I,\,II}_\alpha \simeq \mp 
m^{(D)}_\alpha $ (this case is known as the pseudo--Dirac case). Note that in 
the case of $ m^{(L)}_\alpha \simeq m^{(R)}_\alpha $ the mass neutrinos $ 
\nu^{I}_\alpha $ and $\nu^{II}_\alpha $ are in an obvious analogy to the mesons
$ K_L = p K^0 - q \overline{K^0}$ and $ K_S = q K^0 + p \overline{K^0}$, where 
$ q/p \simeq 1 - 2 \tilde{\varepsilon} \simeq 1 $ is a counterpart of our 
$\tan \theta_\alpha \simeq 1 - \left( m^{(R)}_\alpha - m^{(L)}_\alpha \right)/
2 m^{(D)}_\alpha \simeq 1 $.

 Any model with~~$ m^{(L)}_\alpha\, \simeq \, m^{(R)}_\alpha $, leading to the 
nearly maximal mixing~~$\nu^{I,\,II}_\alpha\, \simeq $ $\left(\nu^{(a)}_\alpha 
\mp \nu^{(s)}_\alpha \right)/\sqrt{2}$, is orthogonal to the popular see--saw 
model with $ m^{(L)}_\alpha \ll m^{(D)}_\alpha \ll m^{(R)}_\alpha $ which gives
$ \nu^{I}_\alpha \simeq \nu^{(a)}_\alpha $ and $\nu^{II}_\alpha \simeq \nu^{(s)
}_\alpha $. In fact, in this case we get from Eqs. (A.5) and (A.7) 

$$
m^{I}_{\alpha} \simeq -\frac{m^{(D)\,2}_\alpha}{m^{(R)}_\alpha} \simeq 0\;\;,\;
\;m^{II}_{\alpha} \simeq m^{(R)}_\alpha + \frac{m^{(D)\,2}_\alpha}{m^{(R)
}_\alpha} \simeq m^{(R)}_\alpha \eqno({\rm A}.8)
$$

\ni and

$$
\cos \theta_\alpha \simeq 1 - \frac{1}{2}\left( \frac{m^{(D)}_\alpha}{m^{(R)
}_\alpha} \right)^2 \simeq 1\;\;,\;\; \sin \theta_\alpha \simeq 
\frac{m^{(D)}_\alpha}{m^{(R)}_\alpha} \simeq 0\;\;.
\eqno({\rm A}.9)
$$

\ni In both cases, however, we may have very small $m^{I}_\alpha $. Notice that
the present experimental limit on the (still not observed) neutrinoless double 
$\beta $ decay (violating the lepton number $ L $) allows for $ m^{(L)}_e $ of 
the order of 1 eV or smaller in both cases of $ m^{(L)}_e \simeq m^{(R)}_e $ 
and $ m^{(L)}_e \ll m^{(R)}_e $.

 With the use of the neutrino mass matrix (A.4) we get the "pure--Majorana" 
oscillation formulae

$$
P\left(\nu^{(a)}_\alpha \rightarrow \nu^{(s)}_\beta \right) = |\langle 
\nu^{(s)}_\beta |e^{-i H t}|\nu^{(a)}_\alpha \rangle |^2 = \delta_{\beta\,
\alpha} \sin^2 2\theta_\alpha \sin^2\left( x^{II}_\alpha - x^{I}_\alpha \right)
\eqno({\rm A}.10)
$$

\ni and

$$
P\left(\nu^{(a)}_\alpha \rightarrow \nu^{(a)}_\beta \right) = |\langle 
\nu^{(a)}_\beta |e^{-i H t}|\nu^{(a)}_\alpha \rangle |^2 = \delta_{\beta\,
\alpha} - P\left(\nu^{(a)}_\alpha \rightarrow \nu^{(s)}_\beta \right)\;\;,
\eqno({\rm A}.11)
$$

\ni where $ x^{I,\,II}_\alpha = 1.27(m^{I,\,II}_\alpha)^2 L/E $ with $ m^{I,\,
II}_\alpha $, $ L $ and $ E $ expressed in~~eV, km and GeV, respectively. Here,
$\sin^2 2\theta_\alpha \simeq 1 $ if $ m^{(L)}_\alpha \simeq m^{(R)}_\alpha $.

 For a form of neutrino mass matrix more general than the "pure--Majorana" form
(A.4), more general mass spectrum and mixing appear. A fairly general mixing 
may be given by the following anticipated formulae for Majorana mass neutrinos:

$$
\nu^{I,\,II}_i = \sum_\alpha U^{(\nu)*}_{\alpha\,i} \nu^{I,\,II}_\alpha =
\sum_\alpha U^{(\nu)*}_{\alpha\,i} \left\{\begin{array}{l} \cos \theta_\alpha\,
\nu_\alpha^{(a)} - \sin \theta_\alpha \,\nu_\alpha^{(s)} \\ \sin \theta_\alpha 
\,\nu_\alpha^{(a)} + \cos \theta_\alpha \,\nu_\alpha^{(s)} \end{array} \right.
\eqno({\rm A}.12)
$$

\ni (with $ i = 1,2,3 $ and $\alpha = e\,,\,\mu\,,\,\tau $). Here $ U^{(\nu)} =
\left( U^{(\nu)}_{\alpha\,i} \right)$ is a $ 3\times 3 $ family unitary matrix
diagonalizing a $ 3\times 3 $ neutrino family mass matrix $ M^{(\nu)} = \left( 
M^{(\nu)}_{\alpha \beta} \right)$ through the relation $\left( U^{(\nu)\,
\dagger} M^{(\nu)} U^{(\nu)}\right)_{ij} = \delta_{ij} m_i $.

 If the Majorana mixing angle $\theta_\alpha $ is taken as a universal $\theta
$ (what certainly would be the case for $\theta_\alpha = 45^\circ $ correspon%
ding to $ m^{(L)}_\alpha = m^{(R)}_\alpha $), then the mixing (A.12) follows 
from the $ 6\times 6 $ neutrino mass matrix

$$
\widehat{M}^{(\nu)} = \left(\widehat{M}^{(\nu)}_{\alpha \beta} \right)\;\;
{\rm with}\;\;\widehat{M}^{(\nu)}_{\alpha \beta} = M^{(\nu)}_{\alpha \beta} 
\left( \begin{array}{cc} \lambda^{(L)} & \lambda^{(D)} \\ \lambda^{(D)} & 
\lambda^{(R)} \end{array} \right)\;\;, \eqno({\rm A}.13)
$$

\ni all entries $\lambda^{(L)}\,,\,\lambda^{(R)}$ and $\lambda^{(D)}$ being 
dimensionless. In fact, such a form leads to the $6\times 6 $ unitary matrix

$$
\widehat{U}^{(\nu)} = \left(\widehat{U}^{(\nu)}_{\alpha\,i} \right)\;\;
{\rm with}\;\;\widehat{U}^{(\nu)}_{\alpha\,i} = U^{(\nu)}_{\alpha\,i} 
\left( \begin{array}{cc} \cos \theta & \sin \theta \\ -\sin \theta & 
\cos \theta \end{array} \right) \eqno({\rm A}.14)
$$

\ni which diagonalizes $\widehat{M}^{(\nu)}$ according to the relation

$$
\left( \widehat{U}^{(\nu)\,\dagger} \widehat{M}^{(\nu)} \widehat{U}^{(\nu)}
\right)_{ij} = \delta_{ij}\left( \begin{array}{cc} m^{I}_i & 0 \\ 0 & m^{II}_i
\end{array} \right)\;\;, \eqno({\rm A}.15)
$$

\ni where

$$
m^{I,\,II}_i\! = \! m_i \lambda^{I,\,II}\;\,{\rm with}\,\;\lambda^{I,\,II}\! = 
\!\frac{\lambda^{(L)}\! +\! \lambda^{(R)}}{2} \mp \sqrt{\left(\frac{\lambda^{(L
)}\! - \!\lambda^{(R)}}{2} \right)^2\! +\!\lambda^{(D)\,2}} \simeq \frac{
\lambda^{(L)}\! +\! \lambda^{(R)}}{2} \mp \lambda^{(D)} \eqno({\rm A}.16)
$$

\ni ($ i = 1,2,3 $) are neutrino masses. The approximate equality in Eq. 
(A.16) is valid for $\lambda^{(L)} \simeq \lambda^{(R)}$. Note that the mass 
matrix (A.13) is the direct product of two matrices ($3\times 3 $ and $ 2\times
2 $) containing separately the family and "Majorana" degrees of freedom. Thus,
also the spectrum (A.16) is multiplicative.

 In the case of neutrino mass matrix (A.13), the "pure--Majorana" oscillation 
formulae (A.11) are extended to the form

\vspace{-0.1cm}

\begin{eqnarray*}
\lefteqn{\!\!\!\!P\left(\nu^{(a)}_\alpha \rightarrow \nu^{(a)}_\beta\right)\! =
\!|\langle \nu^{(a)}_\beta |e^{-i H t}|\nu^{(a)}_\alpha \rangle |^2 = \delta_{
\beta\,\alpha} - \sin^2 2\theta \sum_i |U^{(\nu)}_{\beta\,i}|^2 |U^{(\nu)}_{
\alpha\,i}|^2 \sin^2 \left(x^{II}_i \! -\! x^{I}_i \right) } \\ \!\! & &\!\! - 
4\sum_{j>i} U^{(\nu)\,*}_{\beta\,j} U^{(\nu)}_{\alpha\,j} U^{(\nu)}_{\beta\,i} 
U^{(\nu)\,*}_{\alpha\,i}\left\{\cos^4 \theta \sin^2 \left(x^{I}_i \! -\! x^{I
}_j \right) + \sin^4 \theta \sin^2 \left(x^{II}_j \! -\! x^{II}_i \right)
\right. \\ \!\!&  & \;\;\;\;\;\;\;\;\;\;\;\;\;\;\;\;\;\;\;\;\;\;\;\;\;\;\;\;\;
\;\;\;\;\;\;\;\;\left.+\cos^2 \!\theta \sin^2 \!\theta \left[\sin^2 \left(
x^{II}_j \! -\! x^{I}_i \right) + \sin^2 \left(x^{I}_j \! -\! x^{II}_i\right) 
\right]\right\}
\end{eqnarray*}

\vspace{-1.62cm}

\begin{flushright}
(A.17)
\end{flushright}

\vspace{-0.1cm}

\ni which holds when the quartic products of matrix elements $ U^{(\nu)}_{
\alpha\,i}$ are real. In Eqs. (A.17),  $ x^{I,\,II} = 1.27(m_i^{I,\,II})^2 L/E
$. Here, $\sin^2 2\theta \simeq 1 $ and $\cos^2 \theta \simeq 1/2 \simeq 
\sin^2 \theta $ if $\lambda^{(L)} \simeq \lambda^{(R)}$. 

 The neutrino family mass matrix $ M^{(\nu)} = \left(M^{(\nu)}_{\alpha \beta}
\right)$ may be assumed in the form (1) (with $ f = \nu $). Then, in the case 
of small $\xi = M^{(\nu)}_{33}/| M^{(\nu)}_{12}|$ and $\chi = M^{(\nu)}_{22}/
| M^{(\nu)}_{12}|$, the family unitary matrix $ U^{(\nu)} = \left(U^{(\nu)}_{
\alpha i}\right)$ is given in Eqs. (9). In order to derive from the neutrino 
oscillation formulae (A.17) explicit results, we put $\lambda^{(L)} = 
\lambda^{(R)}\;\left(\equiv \lambda^{(M)}\right) $. In this case, the neutrino 
mass matrix (A.13) has the form

$$
\widehat{M}^{(\nu)} = \left(\widehat{M}^{(\nu)}_{\alpha \beta} \right)\;\;
{\rm with}\;\;\widehat{M}^{(\nu)}_{\alpha \beta} = M^{(\nu)}_{\alpha \beta} 
\left( \begin{array}{cc} \lambda^{(M)} & \lambda^{(D)} \\ \lambda^{(D)} & 
\lambda^{(M)} \end{array} \right)\;\;, \eqno({\rm A}.18)
$$

\ni and the neutrino mass spectrum gives $ m^{I,\,II}_i = m_i\left(\lambda^{(M
)} \mp \lambda^{(D)} \right)$, where $ m_i \equiv m_{\nu_i}$ are determined as
in Eqs. (5) implying $ m_3 \stackrel{>}{\sim} | m_2| \gg m_1 $ ($m_2 = -| m_2 |
$).

 With this mass spectrum, the further discussion depends on the ratio of $
\lambda^{(M)}$ and $\lambda^{(D)}$. We will consider two cases: ({\it i}) $
\lambda^{(M)} = \lambda^{(D)}$ or ({\it ii}) $\lambda^{(M)} \ll \lambda^{(D)}$
(the pseudo--Dirac case). We derive from Eqs. (A.17) and (9) the following 
neutrino oscillation formulae: in the case ({\it i})

\begin{eqnarray*}
P\left(\nu^{(a)}_e \rightarrow \nu^{(a)}_e \right) & = & 1 - \frac{48}{49}
\sin^2\left(1.27\frac{4 m_1^2\lambda^{(D)\,2} L}{E}\right) - \frac{97}{2\cdot
49^2} \;, \\
P\left(\nu^{(a)}_\mu \rightarrow \nu^{(a)}_\mu \right) & = & 1 - \sin^2\left(
1.27 \frac{4 m_2^2\lambda^{(D)\,2} L}{E}\right)\;, \\
P\left(\nu^{(a)}_\mu \rightarrow \nu^{(a)}_e \right) & =  &  \frac{1}{4 \cdot 
49} \sin^2\left(1.27\frac{4 (m_3^2 - m_2^2)\lambda^{(D)\,2} L}{E}\right) 
\end{eqnarray*}

\vspace{-1.62cm}

\begin{flushright}
(A.19)
\end{flushright}

\ni or, in the case ({\it ii})

\begin{eqnarray*}
P\left(\nu^{(a)}_e \rightarrow \nu^{(a)}_e \right)\!\! & = &\!\! 1 - \left(
\frac{48}{49}\right)^2 \sin^2\left(1.27\frac{4 m_1^2\lambda^{(M)}\lambda^{(D)} 
L}{E}\right)- \frac{387}{4\cdot 49^2} \;, \\
P\left(\nu^{(a)}_\mu \rightarrow \nu^{(a)}_\mu \right)\!\! & = &\!\! 1 - \frac{
1}{2}\sin^2\left(1.27 \frac{4 m_2^2\lambda^{(M)}\lambda^{(D)} L}{E}\right) -
\sin^2\left(1.27\frac{4 (m_3^2 - m_2^2)\lambda^{(D)\,2} L}{E}\right)\;, \\
P\left(\nu^{(a)}_\mu\! \rightarrow \nu^{(a)}_e \right)\!\! & =  &\!\!\frac{1}{
49} \sin^2 \left(1.27\frac{4 (m_3^2\! -\! m_2^2)\lambda^{(D)\,2} L}{E}\right) -
\frac{1}{2 \cdot 49} \sin^2\left( 1.27 \frac{4 m_2^2\lambda^{(M)}\lambda^{(D)} 
L}{E}\right)\,, \\ & &
\end{eqnarray*}

\vspace{-1.42cm}

\begin{flushright}
(A.20)
\end{flushright}

\vspace{-0.2cm}

\ni where the $L $'s are three different experimental baselines. In these equa%
tions, the negligible constant terms come out from terms containing $\sin^2 $ 
of large phases averaged over many oscillation lengths determined by the 
leading terms with $\sin^2 $  of small phases. The phases in Eqs. (A.19) and 
(A.20) were calculated in both cases from the relations

\begin{eqnarray*}
(m_j^{I,\,II})^2 - (m_i^{I,\,II})^2 & = & m^2_j\left(\lambda^{(M)} \mp 
\lambda^{(D)}\right)^2 - m^2_i\left(\lambda^{(M)} \mp \lambda^{(D)}\right)^2
\;, \\
(m_j^{II,\,I})^2 - (m_i^{I,\,II})^2 & = & m^2_j\left(\lambda^{(M)} \pm 
\lambda^{(D)}\right)^2 - m^2_i\left(\lambda^{(M)} \mp \lambda^{(D)}\right)^2
\;, 
\end{eqnarray*}

\vspace{-1.62cm}

\begin{flushright}
(A.21)
\end{flushright}

\ni working for $\lambda^{(L)} = \lambda^{(R)}\; \left( \equiv \lambda^{(M)}
\right)$. Note that the second and third Eq. (A.20) are not of the two--flavor
form, in contrast to the second and third Eq. (A.19).

 Comparing two first oscillation formulae (A.19) with the results of solar
and atmospheric neutrino experiments [{\it cf.} Eqs (111) and (114)], 
respectively, we get


$$
\frac{48}{49} \leftrightarrow \sin^2 2\theta_{\rm sol} \sim 0.75 \;\;,\;\; 
4 m_1^2 \lambda^{(D)\,2} \leftrightarrow \Delta m^2_{\rm sol} \sim 6.5
\times 10^{-11}\;{\rm eV}^2 
\eqno({\rm A}.22)
$$

\ni and


$$
1 \leftrightarrow \sin^2 2\theta_{\rm atm} \sim 0.82\;\,{\rm to}\;\,1\;\;,\;\; 
4 m_2^2\lambda^{(D)\,2} \leftrightarrow \Delta m^2_{\rm atm} \sim (0.5\;\,{\rm 
to}\;\,6)\times 10^{-3}\;{\rm eV}^2\;. 
\eqno({\rm A}.23)
$$

\ni Hence, we obtain


$$
\frac{m_1}{|m_2|} \sim (3.61\;\,{\rm to}\;\,1.04)\times 10^{-4}
\eqno({\rm A}.24)
$$

\ni and, due to Eqs. (5),


$$
\xi = (49)^{3/2} \frac{m_1}{|m_2|} \sim (12.4\;\,{\rm to}\;\,3.57)\times 10^{-2}\;,
\eqno({\rm A}.25)
$$


\ni while  $ m_3^2 - m_2^2 = 14[(48/49)\xi + \chi] |M^{(\nu)}_{12}|^2 \sim (
1.80\;\,{\rm to}\;\,0.52)| M^{(\nu)}_{12}|^2 $ with $\chi = \xi/16.848$. This
estimation confirms that $\xi \equiv M^{(\nu)}_{33}/|M^{(\nu)}_{12}|$ and 
$\chi \equiv M^{(\nu)}_{22}/|M^{(\nu)}_{12}|$ are small.

 In contrast to solar and atmospheric results, the LSND result ({\it cf.} Ref. 
[6]), say, $\sin^2 2\theta_{\rm LSND} \sim 0.02 $ and $\Delta m^2_{\rm LSND} 
\sim 0.05\;{\rm eV}^2 $ cannot be explained in the case ({\it i}), since in 
the third Eq. (A.19)


$$
4 (m_3^2 - m_2^2)\lambda^{(D)\,2} \ll 4 m_2^2 \lambda^{(D)\,2} \sim (0.5\;\,
{\rm to}\;\;6)\times 10^{-3}\;{\rm eV}^2 < \Delta m^2_{\rm LSND}
\eqno({\rm A}.26)
$$

\ni for the estimation (A.25) ($ m_3^2 \stackrel{>}{\sim} m_2^2 \simeq 49
|M^{(\nu)}_{12}|$).

 In the case ({\it ii}), however, one may try to compare the third Eq. (A.20) 
with the LSND result getting, say,


$$
\frac{1}{49} \leftrightarrow \sin^2 2\theta_{\rm LSND} \sim 0.02 \;\;,\;\; 
(m_3^2- m_2^2)\lambda^{(D)\,2} \leftrightarrow \Delta m^2_{\rm LSND} \sim 0.05
\;{\rm eV}^2\;. 
\eqno({\rm A}.27)
$$

\ni If in the case ({\it ii}) the relation $ 4 m_2^2\lambda^{(M)}\lambda^{(D)} 
\leftrightarrow \Delta m^2_{\rm atm}$ analogical to (A.23) holds approximately 
[{\it cf.} the second Eq. (A.20)], the comparison with (A.27) gives

$$
\frac{4 m_2^2\lambda^{(M)}}{(m_3^2- m_2^2)\lambda^{(D)}} = \frac{\Delta m^2_{
\rm atm}}{\Delta m^2_{\rm LSND}}  \sim (0.1\;\;{\rm to}\;\;1.2)\times 10^{-1}
\eqno({\rm A}.28)
$$

\ni and

$$
\frac{\lambda^{(M)}}{\lambda^{(D)}} = \frac{1}{14}\left(\frac{48}{49}\xi + 
\chi \right)\frac{\Delta m^2_{\rm atm}}{\Delta m^2_{\rm LSND}}  \sim (9.2\;\;
{\rm to}\;\;2.6)\times 10^{-3}  \frac{\Delta m^2_{\rm atm}}{\Delta m^2_{\rm 
LSND}} \sim (0.92\;\;{\rm to}\;\;3.2)\times 10^{-4}\;,
\eqno({\rm A}.29)
$$

\ni since

$$
\frac{m_3^2 - m_2^2}{m_2^2} = \frac{2}{7}\left(\frac{48}{49}\xi + \chi \right)
\eqno({\rm A}.30)
$$

\ni through Eqs. (5) (in making the estimation (A.29) the value (A.25) was 
used, which holds also in the case ({\it ii}) if $4 m_1^2 \lambda^{(M)}
\lambda^{(D)} \leftrightarrow \Delta m^2_{\rm sol}$). Thus, for the value, (A.%
29) of $\lambda^{(M)}/\lambda^{(D)}$ the third Eq. (A.20) might be consistent 
with the LSND result.

 In conclusion of this Appendix, we can say that a simple neutrino mass matrix 
(A.13), operating with three neutrinos $\nu_e\,,\,\nu_\mu\,,\,\nu_\tau $ only 
and being multiplicative in "Majorana" and family degrees of freedom, is 
consistent in a natural way with solar and atmospheric neutrino experiments, 
but not with the LSND result (that still requires confirmation). Such a con%
sistency of "Majorana" option does not differ much from that based on the 
neutrino mass matrix (82) including two Dirac sterile neutrinos $\enu $ and 
$\mnu $. These conclusions were drawn with the use of our family mass matrix 
(1) (with $ f = \nu $), where the dominance of its off--diagonal elements was 
conjectured. The opposite conjecture of dominance of its diagonal elements 
does not change our conclusions essentially. The nearly bimaximal mixing that 
appears in the $\nu^{(a)}_e \rightarrow \nu^{(a)}_e $ and $\nu^{(a)}_\mu 
\rightarrow \nu^{(a)}_\mu $ oscillation formulae (A.19) is a consequence of 
maximal mixings of $\nu^{(a)}_e $ with $\nu^{(s)}_e $ and $\nu^{(a)}_\mu $ 
with $\nu^{(s)}_\mu $, reflecting the equality $\lambda^{(L)} = \lambda^{(R)}$ 
and so, not holding in the see--saw model corresponding to $\lambda^{(L)} \ll 
\lambda^{(D)} \ll \lambda^{(R)}$. 

 When discussing the Majorana flavor neutrinos $\nu^{(a)}_\alpha $ and $\nu^{
(s)}_\alpha $ ($\alpha = e,\,\mu,\,\tau $), one presumes that the superpositions
(A.3) defining formally these objects are really coherent in processes of elec%
troweak interactions which operate on lefthanded chiral fields $\nu_{\alpha 
L} = \nu^{(a)}_{\alpha L}$, ignoring their righthanded counterparts $
\nu_{\alpha R} = \nu^{(s)}_{\alpha R} $.

 The Dirac part of mass term (A.2) and the kinetic term $\sum_\alpha 
\overline{\nu_\alpha} i\gamma \cdot \partial \nu_\alpha $ can be expressed by 
$\nu_\alpha $ as well as $\nu^{(a)}_\alpha $ and $\nu^{(s)}_\alpha $, {\it 
viz.}


$$
-{\cal L}^{(D)}_{\rm mass} =  \sum_\alpha m^{(D)}_\alpha \overline{\nu_\alpha}
\nu_\alpha = \sum_\alpha m^{(D)}_\alpha \left(\overline{\nu^{(s)}_\alpha} 
\nu^{(a)}_\alpha + \overline{\nu^{(a)}_\alpha}\nu^{(s)}_\alpha \right)
\eqno({\rm A}.31)
$$

\ni and, up to the full divergence $i\partial\cdot \sum_\alpha \overline{
\nu_\alpha} \gamma \nu_\alpha $,


$$
{\cal L}_{\rm kin} = \sum_\alpha \overline{\nu_\alpha} i\gamma \cdot \partial
\nu_\alpha = \frac{1}{2} \sum_\alpha \left(\overline{\nu^{(a)}_\alpha} i\gamma
\cdot \partial \nu^{(a)}_\alpha + \overline{\nu^{(s)}_\alpha} i\gamma\cdot 
\partial \nu^{(s)}_\alpha \right)\;.  
\eqno({\rm A}.32)
$$

\ni Thus, the deciding role in the coherence question is played by the Majorana
part of the mass term (A.2),

\begin{eqnarray*}
\!\!\!-{\cal L}^{(D)}_{\rm mass} & = & \frac{1}{2}\sum_\alpha \left(m^{(L)
}_\alpha \overline{\nu^{(a)}_\alpha}\nu^{(a)}_\alpha + m^{(R)}_\alpha 
\overline{\nu^{(s)}_\alpha}\nu^{(s)}_\alpha \right) \\ 
& = & \frac{1}{2} \sum_\alpha \left\{ m^{(L)}_\alpha \left[
\overline{\left( \nu_{\alpha L} \right)^c} \nu_{\alpha L} + \overline{ 
\nu_{\alpha L} } \left(\nu_{\alpha L}\right)^c \right] + m^{(R)}_\alpha \left[
\overline{\left( \nu_{\alpha R} \right)^c}\nu_{\alpha R} + \overline{ \nu_{
\alpha R} } \left(\nu_{\alpha R}\right)^c \right] \right\}\;,\\ & &
\end{eqnarray*}

\vspace{-1.52cm}

\begin{flushright}
(A.33)
\end{flushright}

\vspace{0.1cm}

\ni which can be presented also in terms of Dirac superpositions $\nu_\alpha =
\nu_{\alpha L} + \nu_{\alpha R}$ and $\nu_\alpha^c =(\nu_{\alpha L})^c + 
(\nu_{\alpha R})^c $, but {\it only} if $ m^{(L)}_\alpha = m^{(R)}_\alpha $. 
Hence, if $ m^{(L)}_\alpha \neq m^{(R)}_\alpha $ (or even if $ m^{(L)}_\alpha 
\simeq m^{(R)}_\alpha $ only approximately), the coherence of Majorana super\-%
positions $\nu^{(a)}_\alpha $ and $\nu^{(s)}_\alpha $ seems to be physically 
preferred over the coherence of Dirac superposition $\nu_\alpha $.

\vfill\eject

~~~~
\vspace{0.6cm}

{\bf References}

\vspace{1.0cm}

{\everypar={\hangindent=0.5truecm}
\parindent=0pt\frenchspacing

{\everypar={\hangindent=0.5truecm}
\parindent=0pt\frenchspacing

~1.~W.~Kr\'{o}likowski, in {\it Spinors, Twistors, Clifford Algebras and 
Quantum Deformations (Proc. 2nd Max Born Symposium 1992)}, eds. Z.~Oziewicz 
{\it et al.}, Kluwer Acad. Press, 1993; {\it Acta Phys. Pol.} {\bf B 27}, 
2121 (1996).

\vspace{0.15cm}

~2.~W. Kr\'{o}likowski,{\it Acta Phys. Pol.} {\bf B 21}, 871 (1990); {\it 
Phys. Rev.} {\bf D 45}, 3222 (1992). 

\vspace{0.15cm}

~3.{\it ~Review of Particle Physics}, {\it Eur. Phys. J.} {\bf C 3}, 1 (1998). 

\vspace{0.15cm}

~4.~Y. Fukuda {\it et al.} (\UK Collaboration), {\it Phys. Rev. Lett.} {\bf 81},
1562 (1998); and references therein. 

\vspace{0.15cm}

~5.~{\it Cf. e.g.}, J.N. Bahcall, P.I. Krastov and A.Y. Smirnov, hep--ph/%
9807216v2.

\vspace{0.15cm}

~6.~C.~Athanassopoulos {\it et al.} (LSND Collaboration), {\it Phys. Rev.}
{\bf C 54}, 2685 (1996); {\it Phys. Rev. Lett.} {\bf 77}, 3082 (1996); nucl--%
ex/9709006.

\vspace{0.15cm}

~7.~M. Appolonio {\it et al.} (CHOOZ Collaboration), {\it Phys. Lett.} {\bf B 
420}, 397 (1998).

\vspace{0.15cm}

~8.~{\it Cf. e.g.}, W. Kr\'{o}likowski, hep--ph/9803323, to appear in {\it 
Nuovo Cimento} {\bf A}.

\vspace{0.15cm}

~9.~{\it Cf. e.g.}, W. Kr\'{o}likowski, hep--ph/9808207, to appear in {\it Acta
Phys. Pol.} {\bf B}; also D.W.~Sciama, astro--ph/9811172.

\vspace{0.15cm}

10.~H.J. Lipkin, hep--ph/9901399.

\vspace{0.15cm}

11.~{\it Cf. e.g.}, P. Kar\"{a}nen, J. Maalampi and J.T. Peltoniemi, 
hep--ph/9901403; and references therein.

\vspace{0.15cm}

12.~Recently, A. Geiser, CERN--EP/98--56, hep--ph/9901433; and references 
therein.

\vfill\eject

\end{document}